\documentclass[
  journal=pasa,
  manuscript=research-paper, 
  year=2024,
  volume=XX,
]{cup-journal}
\usepackage[nolist]{acronym}
\usepackage{microtype}
\usepackage{booktabs}

\begin{acronym}
  \acro{AGN}{Active Galactic Nuclei}
  \acro{ALMA}{Atacama Large Millimeter Array}
  \acro{ATCA}{Australia Telescope Compact Array}
  \acro{ATNF}{Australia Telescope National Facility}
  \acro{ATOA}{Australia Telescope Online Archive}
  \acro{AT20G}{Australia Telescope 20 GHz Survey}
  \acro{ASKAP}{Australian Square Kilometre Array Pathfinder}
  \acro{CASS}{CSIRO Astronomy and Space Science}
  \acro{CABB}{Compact Array Broadband Backend}
  \acro{CSIRO}{Australian Commonwealth Scientific and Industrial Research Organisation}
  \acro{CSS}{Compact Steep Spectrum}
  \acro{CTA}{Cherenkov Telescope Array}
  \acro{EMU}{Evolutionary Map of the Universe}
  \acro{ESP}{Early Science Project}
  \acro{FWHM}{Full Width at Half-Maximum power}
  \acro{GPS}{Gigahertz Peak Spectrum}
  \acro{HFP}{High Frequency Peaker}
  \acro{ISM}{interstellar medium}
  \acro{ICM}{Inter Cloud Medium}
  \acro{IFS}{Intermediate frequencies}
  \acro{LMC}{Large Magellanic Cloud}
  \acro{MCs}{Small \& Large Magellanic Clouds}
  \acro{MCELS}{Magellanic Cloud Emission Line Survey}
  \acro{MIRIAD}[{\tt MIRIAD}]{Multichannel Image Reconstruction, Image Analysis and Display}
  \acro{MOST}{Molonglo Observatory Synthesis Telescope}
  \acro{MRC}{Molonglo Reference Catalogue of Radio Sources}
  \acro{MWA}{Murchison Widefield Array}
  \acro{NRAO}{National Radio Astronomy Observatory}
  \acro{NVSS}{NRAO VLA Sky Survey}
  \acro{OPAL}{Online Proposal Applications \& Links}
  \acro{pc}{parsec\acroextra{: 1 pc $\simeq 3.09 \times 10^{16}$ m}}
  \acro{PN}{Planetary Nebula}
  \acro{PWN}{Pulsar Wind Nebulae}
  \acro{PNe}{Planetary Nebulae}
  \acro{POSSUM}{Polarisation Sky Survey of the Universe's Magnetism}
  \acro{RM}{rotation measure}
  \acro{QSO}{Quasi-Stellar Objects}
  \acro{RFI}{Radio-Frequency Interference}
  \acro{RMS}{Root Mean Squared}
  \acro{SCEM}{School of Computing, Engineering and Mathematics}
  \acro{SGR}{Soft Gamma Repeater}
  \acro{SED}{Spectral Energy Distribution}
  \acro{SFG}{Star Forming Galaxies}
  \acro{SI}[$\alpha$]{Spectral Index\acroextra{, $S \propto \nu^\alpha$}}
  \acro{SKA}{Square Kilometre Array}
  \acro{SMB}{Super Massive Black Hole}
  \acro{SMC}{Small Magellanic Cloud}
  \acro{SN}{supernova}
  \acro{SNRs}{supernova remnants}
  \acro{SNR}{supernova remnant}
  \acro{SNe}{supernovae}
  \acro{SUMSS}{Sydney University Molonglo Sky Survey}
  \acro{WISE}{Wide-Field Infrared Survey Explorer}
  \acro{VLA}{Very Large Array} 
  \acro{VLBI}{Very Long Baseline Interferometry} 
  \acro{MYSO}{Massive Young Stellar Object}
  \acro{HFPs}{High Frequency Peakers}
  \acro{ESP}{Early Science Project}
  \acro{YSOs}{Young Stellar Objects}
  \acro{IRAC}{Infrared Array Camera}
  \acro{BL-Lac}{BL~Lacertae}
  \acro{FSRQs}{Flat Spectrum Radio Quasars}
  \acro{IR}{Infrared}
  \end{acronym}

\def\n49{\mbox{{N\,49}}}
\def\HI{\hbox{H{\sc i}}}
\def\HII{\hbox{H\,{\sc ii}}}
\newcommand{\suzaku}{{\it Suzaku}}
\newcommand{\SII}{[S\,{\sc ii}]}
\newcommand{\OIII}{[O\,{\sc iii}]}

  \title{New radio continuum study of the Large Magellanic Cloud Supernova Remnant N49}
   
  \author{M. Ghavam}
  \affiliation{Western Sydney University, Locked Bag 1797, Penrith, NSW, 2751, Australia}
  \email[M. Ghavam]{18534030@student.westernsydney.edu.au}
  
  \author{M.~D. Filipovi\' c}
  \affiliation{Western Sydney University, Locked Bag 1797, Penrith, NSW, 2751, Australia}
  
  \author{R.~Z. E. Alsaberi}
  \affiliation{Western Sydney University, Locked Bag 1797, Penrith, NSW, 2751, Australia}
  
  \author{L. A. Barnes}
  \affiliation{Western Sydney University, Locked Bag 1797, Penrith, NSW, 2751, Australia}
  
  \author{E. J. Crawford}
  \affiliation{Western Sydney University, Locked Bag 1797, Penrith, NSW, 2751, Australia}
  
  \author{F. Haberl}
  \affiliation{Max-Planck-Institut f{\"u}r extraterrestrische Physik, Gie{\ss}enbachstra{\ss}e, 85748 Garching, Germany}
  
  \author{P. J. Kavanagh}
  \affiliation{School of Cosmic Physics, Dublin Institute for Advanced Studies, 31 Fitzwillam Place, Dublin 2, Ireland}
  
  \author{P. Maggi}
  \affiliation{Universit\'e de Strasbourg, CNRS, Observatoire astronomique de Strasbourg, UMR 7550, F-67000 Strasbourg, France}
  
  \author{J. Payne}
  \affiliation{Western Sydney University, Locked Bag 1797, Penrith, NSW, 2751, Australia}
  
  \author{G.~P.~Rowell}
  \affiliation{School of Physical Sciences, The University of Adelaide, Adelaide 5005, Australia}
  
  \author{H. Sano}
  \affiliation{Faculty of Engineering, Gifu University, 1-1 Yanagido, Gifu 501-1193, Japan}

  \author{M. Sasaki}
  \affiliation{Dr. Karl Remeis Observatory, Erlangen Centre for Astroparticle Physics, Friedrich-Alexander University Erlangen-N\"{u}rnberg, Sternwartstr. 7, 96049 Bamberg, Germany}
  
  \author{N. Rajabpour}
  \affiliation{Western Sydney University, Locked Bag 1797, Penrith, NSW, 2751, Australia}

  \author{N. F. H. Tothill}
  \affiliation{Western Sydney University, Locked Bag 1797, Penrith, NSW, 2751, Australia}
  
  \author{D. Uro\v sevi\' c}
  \affiliation{Department of Astronomy, Faculty of Mathematics, University of Belgrade, Studentski trg 16, 11000 Belgrade, Serbia}
  

  \doi{XXXXXXXXXX}

  \received {dd Mmm YYYY}
  \revised  {dd Mmm YYYY}
  \accepted {dd Mmm YYYY}
  \published{dd Mmm YYYY}

\begin{document}
\begin{abstract}
We present new \ac{ATCA} radio observations toward \n49, one of the brightest extragalactic \acp{SNR} located in the \ac{LMC}.
Our new and archival \ac{ATCA} radio observations were analysed along with {\it Chandra} X-ray data. 
These observations show a prominent  `bullet' shaped feature beyond the southwestern boundary of the \ac{SNR}.  Both X-ray morphology and radio polarisation analysis support a physical connection of this feature to the \ac{SNR}. 
The `bullet' feature's apparent velocity is estimated at $\sim$1300~km~s$^{-1}$, based on its distance  ($\sim$10~pc)  from the remnant's geometric centre and estimated age  ($\sim$7600~yrs). we estimated the radio spectral index, $\alpha= -0.55 \pm 0.03$ which is typical of middle-age \ac{SNRs}. Polarisation maps created for \n49\ show low to moderate levels of mean fractional polarisation estimated at 7$\pm$1\% and 10$\pm$1\% for 5.5 and 9~GHz, respectively. These values are noticeably larger than found in previous studies. Moreover, the mean value for the Faraday rotation of \ac{SNR} \n49\ from combining  CABB data is 212$\pm$65~rad\,m$^{-2}$ and the maximum value of RM is 591$\pm$103~rad\,m$^{-2}$. 

\end{abstract}

\section{INTRODUCTION }
\label{sec:int}

Observations of \ac{SNRs} place important constraints on star and galaxy formation. \ac{SNRs} evolve to be relatively bright at radio wavelengths as charged particles are ejected and accelerated by shock waves interacting with \ac{ISM}. Radio \ac{SNRs} provide a pathway to explore cosmic magnetic fields and test models of galaxy evolution \citep{2017POBeo..96..185F,2021pma..book.....F}. They are characterised by non-thermal radio-continuum emission \citep{Sturner_1997}.

The \ac{LMC} is among the closest face-on galaxies to the Milky Way.
Its distance has been reliably measured at $\sim$50~kpc \citep{2008MNRAS.390.1762D}. It lies away from the Galactic Plane where extinction is minimal, making the \ac{LMC} an excellent research laboratory for investigating objects such as \ac{SNRs}.

In the past decades, hundreds of radio sources have been catalogued in the \ac{LMC} via radio, X-ray, optical, and IR surveys. Most are classified as either \HII\ regions, \ac{PNe} or \ac{SNRs}, appearing at different evolutionary stages \citep{1998A&AS..130..441F,2017ApJS..230....2B,2017MNRAS.468.1794L,2016A&A...585A.162M,2016Ap&SS.361..108L,2015PKAS...30..149B,2015ApJ...799...50L,2009MNRAS.399..769F}.

\n49\ (LHA~120-N\,49; DEM\,L190 or MCSNR\,J0526--6605) was the first extragalactic \ac{SNR} discovered, over half a century ago \citep{1963Natur.199..681M}. It is one of the brightest extragalactic radio \ac{SNRs}. It is located near the northern boundary of the \ac{LMC}; centered on RA~(J2000)~=~05$^h$26$^m$00.1$^s$ and Dec~(J2000)= --66$^{\circ}$05$'$00.4$''$ \citep{2017ApJS..230....2B} and has a diameter of $\sim$18~pc \citep{2008MNRAS.390.1762D,2019Natur.567..200P}. The \citet{1959sdmm.book.....S} age of this young to mid-age \ac{LMC} \ac{SNR} is estimated to be about 7600 years, with explosion energy of  $E_{0}$~$\sim$1.2~$\times$~10$^{51}$\, erg \citep{2017ApJ...837...36L}.

\begin{figure*}
\includegraphics[width=\textwidth,trim=0 0 0 0, clip]{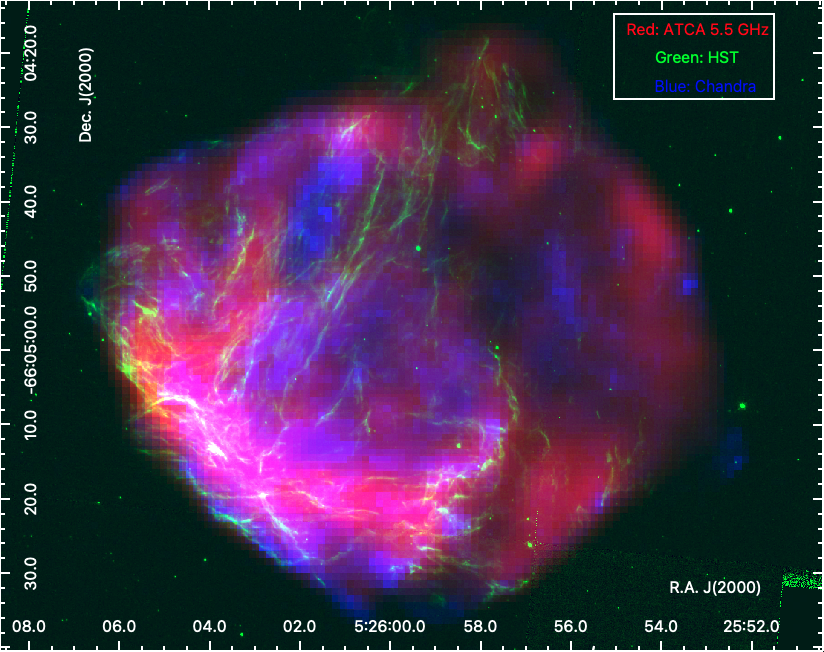}
0\caption{Combined RGB map of \n49\ where Red is radio (5.5~GHz \ac{ATCA}), Green is optical (HST) and Blue is  X-ray (Chandra).}

\label{fig:1}
\end{figure*}

The morphology of \n49\ is somewhat unusual for an \ac{SNR}. Earlier observations show an asymmetric shell; the east southeastern side of the \ac{SNR} is several times brighter than the rest of the object \citep{Vancura1992,2006AJ....132.1877W,1998Dickel}. The most likely explanation for this asymmetry is that the blast wave is expanding into a higher density  \ac{ICM}, causing the shocked material to appear bright in X-ray, UV, and optical wavelengths \citep{Vancura1992} see Figure~\ref{fig:1}.

\n49 is an ideal object for scientific studies because it is bright across multiple wavelength windows including X-ray \citep{2003ApJ...586..210P}, ultraviolet (UV) \citep{Vancura1992}, optical \citep{1992Vancura}, infrared \citep{2006AJ....132.1877W} and radio \citep{1998Dickel}.
Despite having a large amount of data, several key features remain poorly studied. Additionally, their interpretation remains controversial, including a `bullet' feature in the south-west side of the \ac{SNR} and a \ac{SGR} in the north.

%%%%%%%%%%%%%%%%%%%%%%%%%%%%%%%%Fig1%%%%%%%%%%%%%%%%%
%%%%%%%%%%%%%%%%%%%%%%%%%%%%%%%%%%%%%%%%%%%%%%%

Studies using CO data, provide evidence that \n49\ is interacting with a nearby cloud located to the southeast with a tracer of cold molecular hydrogen gas \citep{0004-637X-863-1-55,2023ApJ...958...53S}.  High-excitation optical iron emission lines indicate substantial dust grain destruction occurring in this region \citep{Dopita:2016}. 
Cluster searches in  $\gamma$-ray images reveal high energy photons originating from hadronic processes\footnote{Processes involving particles made up of quarks and subject to the strong force.} through interactions with dense \HII\ regions \citep{2022MNRAS.515.1676Campana}.

{\it Chandra} observations show metal-rich ejecta at both the centre and periphery of  \n49\ \citep{2003ApJ...586..210P,0004-637X-748-2-117}. Beyond the southwestern boundary and towards the eastern boundary are two metal-rich `bullet` features (also known in the literature as shrapnel or knots)  detected in earlier radio and X-ray studies \citep{1998Dickel}. The southwestern bullet is an extended feature with a head ($\sim$4$''$ in radius or 23~pc at  50~kpc) and a tail that connects to the main \ac{SNR} shell (Figures~\ref{fig:4}). Spectral analysis of the head region indicates  Silicon (Si) and Sulphur (S) enhancement most likely caused by multiple evolutionary phases. The enhanced Si lines in the eastern boundary suggest another ejecta-rich region  \citep{0004-637X-748-2-117}. 
In general, earlier morphological studies based on radio data \citep[e.g:][]{1998Dickel}, show similarity to later X-ray studies \citep{2003ApJ...586..210P}. 

There has been controversy regarding \n49's thermonuclear versus core-collapse origin. \cite{0004-637X-748-2-117} argued that the observed Si/S ratio in the ejecta suggests a thermonuclear (type~Ia) origin. Other results seem to imply a core-collapse origin. For example, the recombining plasma (RP) state of \n49  determined from \suzaku\ observations appears consistent with a core-collapse model from a massive star  \citep{0004-637X-808-1-77}.  In this scenario, shocks travelled through a dense circumstellar medium created by a massive progenitor. The high radio brightness of \n49\ also suggests a core-collapse origin, since the radio luminosity of a thermonuclear \ac{SNR} is somewhat less \citep[][their fig.~16]{2017ApJS..230....2B}.

X-ray observations show that the point source \ac{SGR} 0526$-$66 (PSR B0525$-$66) is located in the middle of the northern part of \n49. This particular \ac{SGR} is only one of two known magnetars in the \ac{LMC} \citep{2004AdSpR..33..409P,2009ApJ...700..727B,2023MNRAS.524.5566I}. The position of this \ac{SGR} is RA~(J2000)~=~05$^h$26$^m$00.7$^s$ and Dec~(J2000)= $-$66$^{\circ}$04$'$35.0$''$ and its age has been estimated to $\sim$1900~yrs \citep{0004-637X-748-2-117,2006VInk}. The close proximity of \ac{SGR} 0526$-$66 would normally be sufficient evidence to classify \n49\ as a core-collapse \ac{SNR} \citep{2009ApJ...700..727B}. However, stars more massive than 21.5~M$_\odot$ are very scarce near  \n49, implying that either SGR~0526$-$66 came from a progenitor with a mass below 30~M$_\odot$ \citep{1995HIll} or that the magnetar is merely a chance projection along the line of sight. Support for the latter is found in \cite{1995ApJ...448..623D}. They note the location and spectral indices of the X-ray hot spots of \n49 and SGR\,0526$-$66 are probably not related. Further, \cite{2001ApJ...559..963G} use radio observations of \n49\ and \ac{SGR} 0526$-$66 to conclude that existing claims of associations between the two are implausible.

Motivated by previous studies and a desire to understand \n49's\ features, we obtained new \ac{ATCA} radio observations at 2.1, 5.5, and 9~GHz  
to create higher resolution and more sensitive images.  We also use this data to construct polarisation maps to understand its magnetic fields better. Data and reduction techniques are presented in Section~\ref{data}, results in Section~\ref{results}, and a discussion in Section~\ref{disc}. Finally, conclusions and future prospects are presented in Section~\ref{conclusion}.

\section {Observational Data and Reductions}
 \label{data}

\subsection{Radio data}

Our new \ac{ATCA} observations of \n49  were performed in 30$^\mathrm{th}$ November 2019, 2$^\mathrm{nd}$ January, 10$^\mathrm{th}$ February 2020 and 23$^\mathrm{rd}$ February 2020 (project codes C3275, CX454, CX459 and C3292). These observations utilised a frequency switching mode (between 2100~MHz and 5500 / 9000~MHz) in  6A, 1.5C, and EW367 configurations. Observations were carried out in `snap-shot' mode, with 1-hour integrations over a 12-hour period minimum, using the Compact Array Broadband Backend (CABB) (2048\, MHz bandwidth)  centred at wavelengths of 3/6\,cm \footnote{$\nu$~=~4500--6500 and 8000--10000\, MHz; centred at 5500 and 9000\,MHz, respectively} and 13\,cm ($\nu$~=~2100\,MHz). 

Our data reductions also used available \ac{ATCA} archival observations from the \ac{ATNF} online archive\footnote{https://atoa.atnf.csiro.au}. Two different data sets were combined for analysis. The first was before the CABB upgrade from project C177 (pre-CABB; 1990s, with 128-MHz bandwidth) and included observations at \ac{IFS} 2.4, 4.5, 4.79, 6 and 8.64~GHz \citep{1995Dickel,1998Dickel}. As the remnant is so evolved combining data from 25 years will not introduce errors. The second data set, C520 (pre-CABB), includes observations at \ac{IFS} 4.79 and 8.64~GHz \citep{1998MNRAS.298..692F}. Each 128 MHz--wide frequency band was correlated in 33 separate channels of equal width. Details of these observations are listed in Table~\ref{tab:1}.

%%%%%%%%%%%%%%%%%%         table 1    %%%%%%%%%%%%%%%%%%%%%%%%%%%%%%%
\begin{table*}[h!]

	\caption{Details of \ac{ATCA} Observations of \ac{SNR} \n49 used in this study. Source 1934--638 was used as a primary (bandpass and flux density) calibrator in all observations.}
	\label{tab:1}
	\begin{tabular}{lccccccl} 
		\hline\hline
Observing & Project  & Frequency & Array     & Integration Time  & Bandwidth & Phase& Reference \\
Date      & Code     & (GHz)     & Config.   & (minutes)         & (MHz)   & Calibrator &      \\
            \hline
1992 Apr 23 & C177 & 2.4       & 6C  &523.3 & 128 &0407--658& \cite{1995Dickel} \\
1992 Jun 5  & C177 & 2.4       & 1.5D&401.1 &128 &0407--658& \cite{1995Dickel} \\
1992 Jun 17 & C177 & 2.4       & 1.5B&347 & 128 &0407--658& \cite{1995Dickel} \\
1993 Jul 19 & C177 & 4.5,6     & 6C &361.2  & 128&0454--810& \cite{1998Dickel} \\
1993 Aug 27 & C177 & 4.79, 8.64 & 1.5B& 528.2& 128&0454--810 &\cite{1998Dickel} \\
1993 Oct 12 & C177 & 4.79, 8.64 & 1.5D& 700.5& 128&0454--810& \cite{1998Dickel} \\
1994 Oct 20 & C177 & 4.79, 8.64 & 750C&719.2 & 128&0454--810 & \cite{1998Dickel}\\
1994 Nov 26 & C177 & 4.79, 8.64 & 375&607.7  & 128 &0454--810& \cite{1998Dickel} \\
1996 Jul 5  & C520 & 4.8, 8.64  & 6C & 483.0 & 128 &0515--674& \cite{1998MNRAS.298..692F} \\
1996 Jul 7  & C520 & 4.8, 8.64  & 6C &106  & 128 &0515--674 & \cite{1998MNRAS.298..692F}\\

2019 Nov 30 & C3275 & 2.1, 5.5, 9  & 1.5C &21.6  & 2048 &0530--727 & This work\\
2020 Jan 2  & CX454&5.5, 9 &1.5C&201& 2048 &0530--727 & This work\\

2020 Feb 10 & CX459&5.5, 9&6A &198& 2048 &0530--727 & This work\\
2020 Feb 23 & C3292&2.1, 5.5, 9&EW367 &150& 2048 &0530--727 & This work\\

		\hline
	\end{tabular}
\end{table*}
%%%%%%%%%%%%%%%%%%%%%%%%%%%%%%%%%%%%%%%%%%%%%%%%%%%%%%%%%%%%%%%%%%%%%

Data reduction and imaging were performed using the \textsc{miriad}\footnote{http://www.atnf.csiro.au/computing/software/miriad/}  \citep{1995ASPC...77..433S}, \textsc{karma}\footnote{http://www.atnf.csiro.au/computing/software/karma/} \citep{1995Gooch} and \textsc{DS9}\footnote{\url{https://sites.google.com/cfa.harvard.edu/saoimageds9}} \citep{2003ASPC..295..489J} packages. We measured \n49\ flux densities (Table~\ref{tab:2}) as  described in \citet[][Section 2.4]{2019PASA...36...48H} and \citet[][]{2022MNRAS.512..265F,2023MNRAS.518.2574B} .

%%%%%%%%%%% Table 2    %%%%%%%%%%%%%%%%%%%%%%%%%%%%%%%
\begin{table}[h!]
	\centering
	\caption{Details of CABB and pre-CABB radio continuum \n49\ images created and used in this study.}
	\label{tab:2}
	\begin{tabular}{lcccl} 
		\hline\hline
      Frequency  & Beam Size                    & RMS                 &  S$_{\nu}$& Note    \\
	      (GHz)  & (arcsec)                     &  (mJy~beam$^{-1}$)  & (Jy)          \\
		    
		    \hline
		    2.1  & 4$''$ $\times$ 4$''$ & 0.128 & 1.09& CABB \\
		    2.4  & 8.9$''$ $\times$ 6.6$''$ & 0.100 & 0.930 &pre-CABB\\
		    4.79 & 5$''$ $\times$ 5$''$     & 0.067 & 0.838 &pre-CABB\\
		    5.5  & 5$''$ $\times$ 5$''$ & 0.055 & 1.02&CABB \\
		    8.64 & 5$''$ $\times$ 5$''$     & 0.042 & 0.651&pre-CABB \\
		    9.0  & 4$''$ $\times$ 4$''$ & 0.064 & 0.30&CABB \\
		    \hline
\end{tabular}	
\end{table}
%%%%%%%%%%%%%%%%%%%%%%%%%%%%%%%%%%%%%%%%%%%%%%%%%%%%%%%%%%%%%%%%  

To obtain the 5.5 and 9~GHz frequency images, four sets (`12-h days') of observational data (projects C3275, CX454, CX459, and C3292) were combined. 
In all cases, 1934--638 was used as the standard bandpass and flux density calibrator while the phase calibrator was 0530--727. Automated task \textsc{pgflag} was used to flag errors in the data;  tasks \textsc{blflag} and \textsc{uvflag} were used for minor manual flagging. A robust weighting of 0.5 created the best radio images at 5.5 and 9~GHz.

\begin{figure*}
 \includegraphics[width=\textwidth,trim=0 0 10 0, clip]{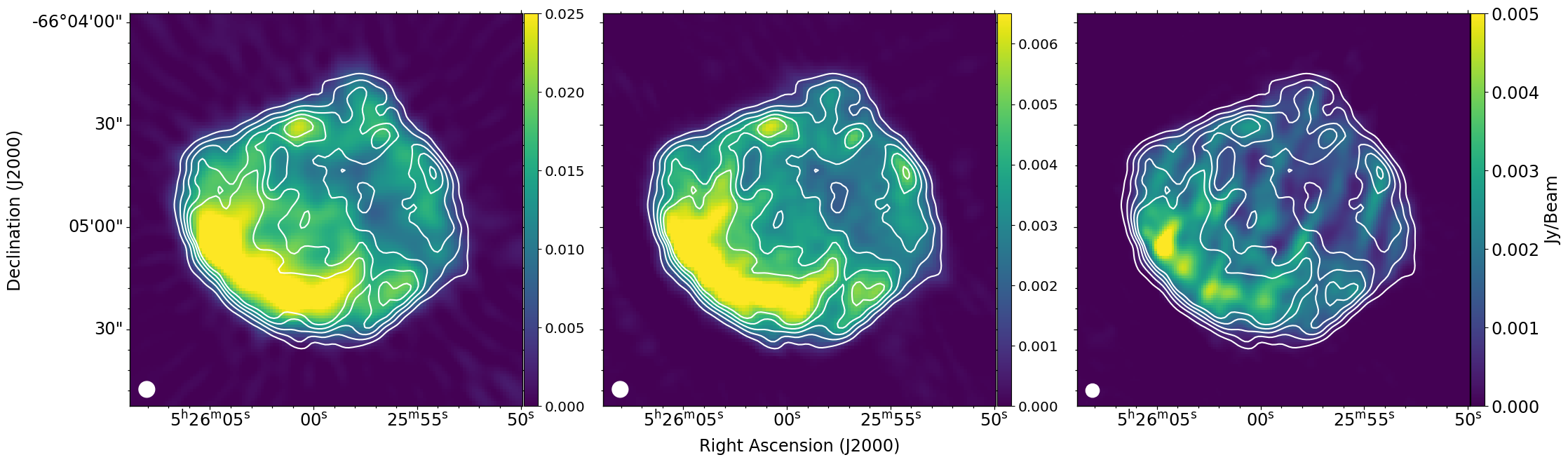}

  \caption{CABB radio continuum images  of \ac{SNR} \n49.
  5.5 GHz contour levels overlaid on each of the 3 images are 1, 2, 3, 4, and 5~mJy~beam$^{-1}$. 
  {\bf Left:} 2.1 GHz image  
  with a resolution (beam size) of 4$''$ $\times$ 4$''$ 
  {\bf Middle:} 5.5 GHz 
   image with a resolution (beam size) of 5$''$ $\times$ 5$''$ 
  {\bf Right:}  9 GHz image with a resolution (beam size) of 4$''$ $\times$ 4$''$. White circles in the lower left corner scale the synthesised beam.}
  \label{fig2}
\end{figure*}

Likewise, to create the 2.1~GHz  image, two observational sets (days) of data were combined from projects C3275 and C3292. Two array configurations were available: 1.5C and 6A. Imaging processing was similar to that used above for the 5.5 and 9~GHz images. Figure~\ref{fig2} shows total intensity images at 2.1, 5.5, and 9~GHz, respectively.

%%%%%%%%%%%%%%%%%%%%%%%%%%%%%%%%%%fig 2 %%%%%%%%%%%%%%%%%%%%%%%%%
%%%%%%%%%%%%%%%%%%%%%%%%%%%%%%%%%%%%%%%%%%%%%%%%%%%%%%%%%%%%%%%%%%%%%%%%%%

To compare new CABB data with pre-CABB upgrade observations, three sets of pre-CABB data (project C177) were combined to obtain a 2.4~GHz frequency image of \n49. Array configurations at this frequency included 1.5B, 1.5D, and 6C. 
Again, a robust weighting of 0.5 created the best radio images. 
The \textsc{miriad} task \textsc{selfcal} was also applied to perform self-calibration of the visibility data \citep{1995AAS...18711202B}. Phase \textsc{selfcal} was performed 3 times which improved image resolution compared to the previous work of \citet{1995Dickel}.

%%%%%%%%%%%%%%%%%%%%%%%%%%%%%%%%%%fig 3 %%%%%%%%%%%%%%%%%%%%%%%%%%%%%%%%%%
\begin{figure*}
\centering
 \includegraphics[width=\textwidth,trim=0 0 0 0, clip]{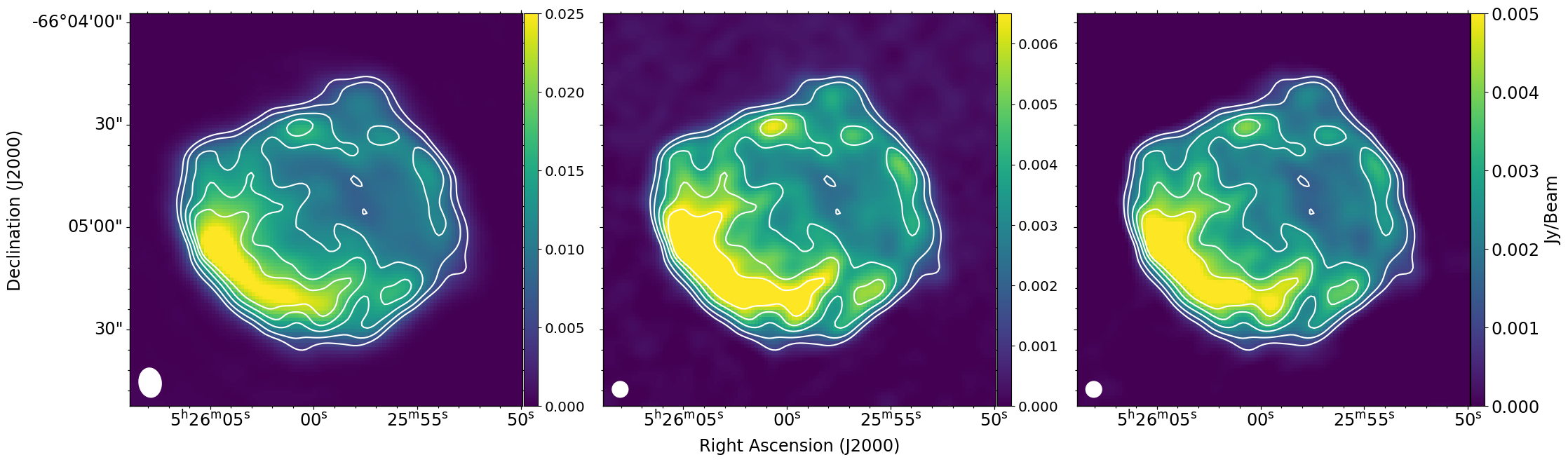}
  \caption{Pre-CABB update radio continuum images of \ac{SNR} \n49. 4.79 GHz contour levels overlaid on each of the 3 images are 0.85, 1.5, 2.5, 3.5, and 4.5~mJy~beam$^{-1}$. {\bf Left:} 2.4 GHz image with a resolution (beam size) of 8.9$''$ $\times$ 6.6$''$. {\bf Middle:} 4.79 GHz image with a resolution (beam size) of 5$''$ $\times$ 5$''$. {\bf Right:} 8.64 GHz image with a resolution (beam size) of 5$''$ $\times$ 5$''$.  White circles in the lower left corner scale the synthesised beam. }
  \label{fig3} 
\end{figure*}
%%%%%%%%%%%%%%%%%%%%%%%%%%%%%%%%%%%%%%%%%%%%%%%%%%%%%%%%%%%%%%%%%%%%%%%%%%

To obtain 4.79 and 8.64~GHz images, data from projects C177 and C520  utilising configurations 1.5B, 1.5D, 6C, 750C, and 375 were all combined for their respective frequencies. 
A robust weighting of --0.5 was used for 4.79~GHz and 0.5 for 8.64~GHz image reduction. The resolution of both 4.79 and 8.64~GHz images were smoothed to 5$''$ $\times$ 5$''$ using the \textsc{miriad} task \textsc{invert}. Tasks \textsc{pmosmem} and \textsc{restor} created the resulting convolved mosaic images. Integrated \n49\ flux densities at 4.79 and 8.64~GHz are 0.838 $\pm$  0.1 ~Jy and 0.651 $\pm$ 0.1 ~Jy, respectively. Most of the uncertainties reported are systematic due to calibration and statistical fluctuations.

Figure~\ref{fig3} shows the total intensity of pre-CABB images at 2.4, 4.79, and 8.64~GHz. Details of the images used in this study are listed in Table~\ref{tab:2}.

\subsubsection{Polarimetry}
\ac{SNR} polarimetry can provide details on the intensity of the polarised emission, polarisation degree, and magnetic field orientation. Moreover, \ac{SNRs} magnetic fields play a significant role in the cosmic ray acceleration mechanism \citep{doi:10.1093/mnras/sty751}. By combining radio with X-ray data, the strength of the actual magnetic field can be determined \citep{2011G.Dubner}.
    
Polarisation maps can be created from full Stokes parameters, including $I$, $Q$, $U$, and $V$ data that describe multiple wave components \citet{2018Robishaw} and \citep{trove.nla.gov.au/work/14706507}.

Formally, the linear polarisation state of a radio wave as an observable quantity can be described as 
\begin{equation}
    \textbf{P}=Q + iU = pIe^{2i\chi}
    \label{eq1}
\end{equation}
where $I$ is the flux density of the radio emission (in units of Jy~beam$^{-1}$), $p$ the linear fractional polarisation, and $\chi$  the observed position angle of the radio wave given by
\begin{equation}
\chi=\frac{1}{2}\tan^{-1}\left(\frac{U}{Q}\right).~
\label{eq2}
\end{equation}

Our polarised intensity maps produced by the vector combination of the Stokes $Q$ and $U$ maps were statistically corrected for bias using the \textsc{miriad} task \textsc{impol}. The mean fractional polarisation has been evaluated in two ways: (1) by dividing the polarised intensity map by the total intensity map, determined as a mean ratio, and (2) by dividing the integrated polarised intensity of the source by the integrated total intensity.\textsc{impol} was used to calculate the Stokes $I$, $Q$, and $U$ parameters, from which were derived the polarised emission, position angle, and resultant uncertainty maps (Equations~\ref{eq1} and ~\ref{eq2}).

Whenever an electromagnetic wave passes through magnetized plasma or magneto--ionic material, the polarisation angle of the wave is rotated from its intrinsic value ($\chi_{0}$), due to the different phase velocities of the two opposite-hand polarisation modes. This effect is known as Faraday rotation.
For an astronomical radio source, the total observed Faraday rotation along the line of sight can be estimated from the variation of the position angle with respect to the observing wavelength, ${\Delta\chi}/{\Delta\lambda_{obs}^2}$. This quantity is known as \ac{RM}.

The measured change in position angle of radiation, ${\Delta\chi}$, for an electromagnetic wave arriving at an observer, assuming a single magnetic field with only one source along the line of sight and no internal Faraday rotation or beam depolarisation is defined as:
 \begin{equation}
\Delta\chi=\chi_{obs}-\chi_{0}=RM\lambda_{obs}^{2},
\end{equation}
where $\chi_{obs}$ and $\chi_{0}$ (rotation angle of the source) are in radians and the wavelength of the observed radiation, $\lambda_{obs}$, in meters.

Using the assumptions above and the equation for Faraday depth, the \ac{RM} (rad~m$^{-2}$) can be normalized to  physical parameters:
\begin{equation}
RM=0.81\int_{0(source)}^{L(observer)} n_{e}B_{||}dl~~ ,
\label{eq3}
\end{equation}
where $n_{e}$ is the number density of thermal electrons (in units of cm$^{-3}$), $B_{||}$ is the magnetic field component in the direction of the line of sight to the source (in $\mu$G) and $l$ is the displacement (in pc) between the source and observer. Equation~\ref{eq3} indicates that the \ac{RM} represents a relation between the observed position angle, $\chi_{obs}$, and the magneto-ionic properties of plasma in radio sources. \citep{1966MNRAS.133...67B}.

To determine the \ac{RM} and the true orientation of the magnetic field (B), observations at two or more frequencies are required. This is to resolve a degeneracy of the unknown number of half cycles of rotations of the position angle between the source and observer; i.e., the measurements of the position angles are uncertain by $\pm n\pi$. More details on how to remove this ambiguity and determine an accurate rotation measure by careful selection of observational wavelengths can be found in \citep{2004Clarke...37..337C}. However, one still can argue that the entire source cannot be described by a single \ac{RM} \citep{2012MNRAS.421.3300O}.

The \textsc{miriad} task \textsc{imrm} fits the position angle image vectors from at least two position angle images up to five different frequencies on a pixel-by-pixel basis. In the same process, the task computes the unrotated angle $\chi_{0}$. In this study, we used combined images at 5.5 and 9~GHz (CABB images) to obtain a reasonable \ac{RM} estimate. 

There was no reliable evidence of polarisation emission at 2.1~GHz image, most likely, because the depolarization at these frequencies is caused by Faraday rotation.
\textsc{miriad} task \textsc{uvsplit} was used to break these wide-band frequencies (2.048~GHz) into two narrower bands, each with a maximum bandwidth of 1 GHz. The cumulative effect of these four 
smaller frequencies can help identify the overall trend more accurately and reduce the ambiguity in the RM calculation.
While using only two broadband frequencies leads to ambiguities due to the periodic nature of the polarization angle, splitting these into smaller frequencies can help mitigate but not fully resolve these ambiguities. Each additional frequency provides more data points, improving the fit and potentially identifying the correct RM. 
In all four bands, we only calculated \ac{RM} when the signal-to-noise ratio was at least 10$\sigma$. Furthermore, $\sim$10\% error is due to the $n~\pi$ ambiguity in measurements of the position angle \citep{2005A&A...441.1217B} as we use four bands in our \ac{RM} estimation.

\subsection{X-Ray data} 

\begin{figure*}
\includegraphics[width=0.9\textwidth,trim=0 0 0 0,clip]{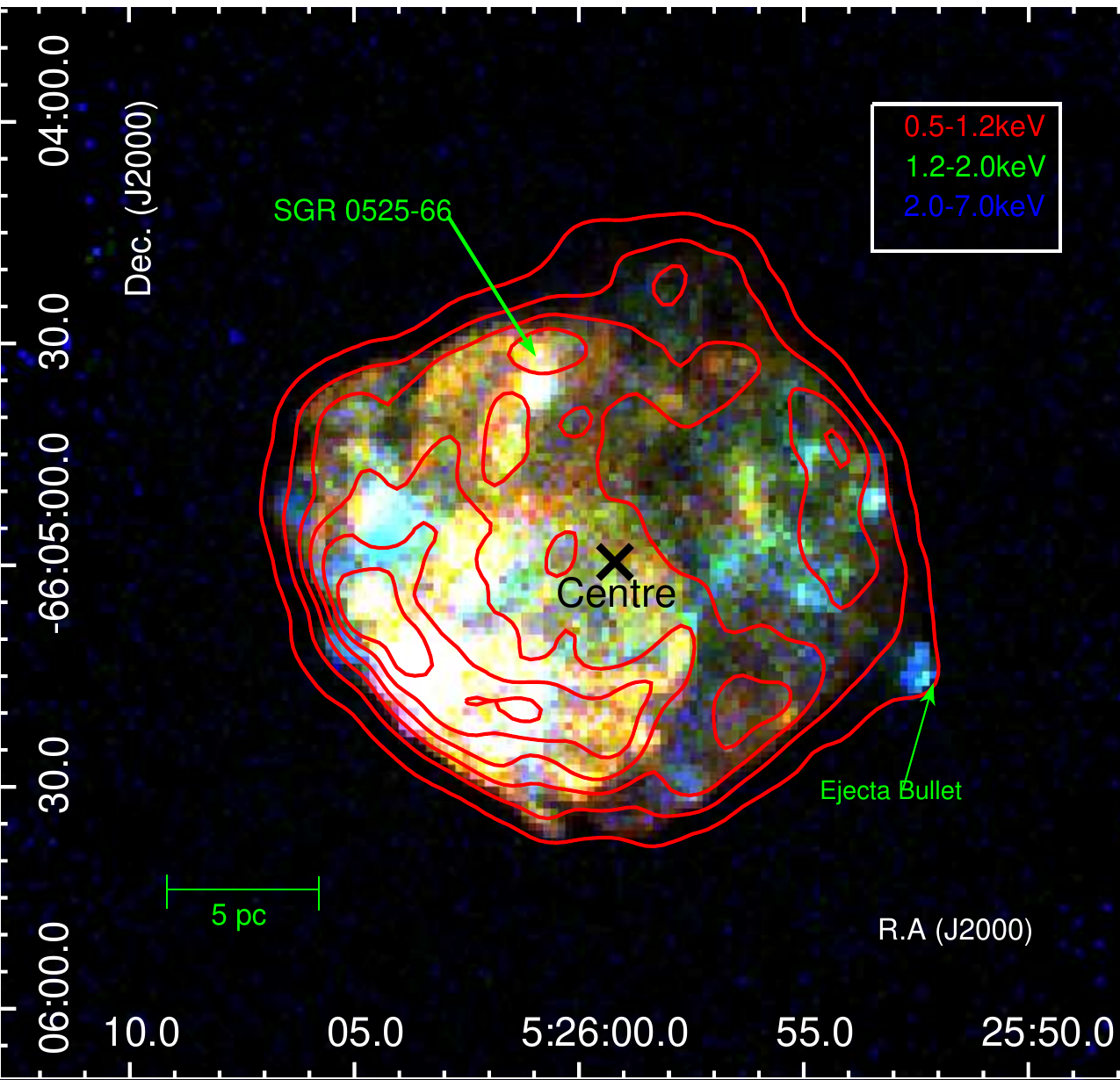}
\caption{An X-ray three-colour image of \n49\ that shows our new Radio data superimposed as contours. Red represents the energy band 0.5--1.2~keV (soft), green corresponds to the energy band 1.2--2.0~keV (Medium) and blue represents the energy band 2.0--7.0~keV (hard). This X-ray image is overlaid with 5.5~GHz radio contours at 0.25, 1.4, 2.6, 3.8, and 5~mJy\,beam$^{-1}$. The distance between \ac{SNR} centre to the `bullet' feature is used to estimate the upper and lower angular displacement values as explained in Table~\ref{tab:3}}. 
    \label{fig:4}
\end{figure*}

\citet{0004-637X-748-2-117} presented an X-ray study on \n49\ using the Advanced CCD Imaging Spectrometer (ACIS) on board the {\it Chandra} X-Ray Observatory. The \ac{SNR} was observed on 18$^\mathrm{th}$~July~2009 and 19$^\mathrm{th}$~September~2009 (ObsIDs 1023, 10806, 10807 and 10808) during AO10. To compare these with our new radio data, we re-processed the original X-ray data using the CIAO v4.9 software package with CALDB v4.7.6 \citep{2006F}.  The merge-obs\footnote{http://cxc.harvard.edu/ciao/ahelp/merge\textunderscore obs.html}  and chandra-repro\footnote{http://cxc.harvard.edu/ciao/ahelp/chandra\textunderscore repro.html} scripts  \citep{2006F} were used to create the combined energy-filtered and exposure-corrected images. The total effective exposure time after data reduction is $\sim$108\,ksec which matches the effective exposure time of the original reduced image stated in \cite{0004-637X-748-2-117}. The re-reduced three-colour (RGB) X-ray image of the soft (0.5--1.2~keV), medium (1.2--2.0~keV), and hard (2.0--7.0~keV) energy bands is presented in Figure~\ref{fig:4}.

%%%%%%%%%%%%%%%%%%%%%%%%%%% fig4
%%%%%%%%%%%%%%%%%%%%%%%%%%%%%%%%%%%%%%%%%%%%%%%%%%%%%%%%

\subsection{Optical data} 

Optical HST images (H$\alpha$, \SII, and \OIII) were downloaded from the Mikulski Archive for Space Telescopes\footnote{\url{https://mast.stsci.edu/portal/Mashup/Clients/Mast/Portal.html}}. All information regarding these HST observations and their data reduction are given in \cite{2007AJ.Bilikova}. These data have been used to create our optical images in this study.

\section{Results and Analysis}
\label{results}

\subsection{Morphology}

Figure~\ref{fig:4} shows the three-colour {\it Chandra} ACIS image of \n49\ where complex and asymmetric X-ray emission features are evident. 

 It is evident from the RGB image that the softest X-rays are significantly inside the SNR. The outer boundary of the \n49\ shell exhibits extensive regions of green and blue, which are notably harder. In most cases, the outer shock tends to be harder compared to the reverse shock within the ejected material \citep{Leahy_2020}. Two metal-rich ejecta as suggested by \citet{0004-637X-748-2-117} have been detected: one towards the southwest and another, less obvious one (not studied here) 
towards the east to northeast. The western (to southwestern) half of \n49\ is dominated by hard X-rays, which indicates the blast wave is expanding into low-density \ac{ISM}.  Our radio data also suggest this. 
\cite{0004-637X-808-1-77} noted that the X-ray images of \n49 show an irregular morphology that is brightest inside the radio shell, as is a typical characteristic of a mixed morphology \ac{SNR}. 
Also, optical images suggest an outflow pointing up to the north direction, which could be indicative of some asymmetry in the progenitor SN event.

In Figure~\ref{fig:4}, Using a  method by \footnote{\url{https://github.com/ccollischon/banana}}  \citet{2021A&A...653A..16C}, we estimated the geometrical center of the \n49\ radio image as RA~(J2000)~=~05$^h$25$^m$59.2$^s$ and Dec~(J2000)= --66$^{\circ}$04$'$59.6$''$. We note there is a soft gamma repeater,   \ac{SGR} 0526--66,   positioned at RA~(J2000)~=~05$^h$26$^m$00.9$^s$ and Dec~(J2000)= --66$^{\circ}$04$'$31.6$''$.

The hard X-ray knot in the southwestern part of \n49, beyond the remnant's shock boundary has been called the `bullet' and is located at RA~(J2000)~=~05$^h$25$^m$52.4$^s$ and Dec~(J2000)= 
--66$^{\circ}$05$'$14.7$''$. This feature has emission lines that dominate the knot's X-ray spectrum which can be modeled by a metal-rich hot thermal plasma \citep{2003Park}. Based on this result, it appears it was created by ejected fragments from the SN explosion that produced \n49. Additionally,  there is evidence of a weak trail behind the head of the bullet \citep{2003ApJ...586..210P}.

\citet{2003ApJ...586..210P} notes strong Si enhancement in the eastern region of the remnant corresponding to an internal cloud(area of a low density region between clouds of interstellar material within \n49) which resembles the spectral properties of the head of the `bullet'. This region may be a candidate ejecta feature (positioned at RA~(J2000)~=~05$^h$25$^m$52.2$^s$ and Dec~(J2000)= --66$^{\circ}$05$'$16.7$''$), probably isolated from the complex emission due to shocked dense clumpy clouds \citep{2003ApJ...586..210P}.

\subsubsection{The `Bullet's' Muzzle (Kinetic) Energy }

\n49\ exhibits  features rarely seen in \ac{SNRs} of similar age and type. Most prominent is the X-ray ejecta `bullet'  (Figure~\ref{fig:4}) extending beyond the remnant's shell in the southwest.  This feature has predominantly hard emission (2.0--7.0~keV), as shown in blue. Outer shell-like X-ray features in \n49\ are largely coincident with radio shell contours as seen in Figures~\ref{fig:4}, most likely indicating regions of blast wave shock. In fact, hard X-ray emission is reasonably correlated with radio and appears to peak toward the southeast edge. 

The average velocity of the `bullet' has been estimated at $\sim$1300~km~s$^{-1}$, assuming its distance to the \ac{SNR} centre of $\sim$10~pc and remnant age of 7600~yrs \citep{2017ApJ...837...36L}. 
The `bullet's displacement and the measured angle indicate how far and in which direction the \ac{SNR}'s centre might have shifted over time. In this case, we note the SN explosion centre of \n49\, inferred from the projection of the southwest `bullet' trail is shifted the geometrical centre by $\sim$10.4~pc, and forming an angle of $\sim$75$^{\circ}$with the initial centre.

The high temperature and large ionisation timescale of this bright X-ray knot support its origin as a high-speed (bow-shock) ejecta. 
As the \ac{SNR} encounters regions of varying density in the \ac{ISM}, these interactions contribute to the asymmetrical shape that leads to a bow shock. After the \ac{SNR} expands and interacts with the surrounding \ac{ISM}, influenced by factors such as relative motion and \ac{ISM} density variations, a bow shock can eventually develop.

In an \ac{SNR} with a density gradient in the surrounding \ac{ISM}, the centre of the remnant is positioned toward the high-density emission. This occurs because the shock wave decelerates more in the denser region, leading to an asymmetrical expansion and accumulation of material on the high-density side. The side facing the denser medium appears more compressed, while the side facing the lower-density medium is more extended, all resulting in the different geometric and mass centres of the \ac{SNR} \citep{2012PTEP.2012aA309J}.

%%%%%%%%%%%%%%%%%%%%%%% Table 3 %%%%%%%%%%%%%%%%%%%%%%%
\begin{table}
	\centering
	\caption{\n49\ kick velocity estimate. Upper and lower angular displacement values were estimated using the \ac{SNR} centre as an upper limit, and the `midpoint' as the lower limit (see Figure~\ref{fig:4}). We assume a distance to the \ac{LMC} of 50~kpc to calculate the physical displacement and an \ac{SNR} age of 7600~yrs to calculate the kick velocity.
}
	\begin{tabular}{lcccc} 
		\hline\hline
 Object  &   Measured angle    & Displacement & Kick Velocity   \\
	     & (degree)            & (parsec)    &  (km~s$^{-1}$) \\
		    
		    \hline
		  Bullet& 75$^{+0.97}_{-0.97}$ & 10.4$^{+0.53}_{-0.53}$ & 1300$^{+66.4}_{-66.4}$ \\
		    \hline
\end{tabular}
\label{tab:3}
\end{table}
%%%%%%%%%%%%%%%%%%%%%%%%%%%%%%%%%%%%%%%%%%%%%%%%%%%%%%%%%%%%%%%%%

To constrain the knot's precise origin is difficult since the original \ac{SN} might neither be in the geometrical centre (although still possible) nor at the ``midpoint''. Bearing this in mind, we estimate a somewhat crude `bullet' kinetic energy, $E_K$, assuming 
a pre-shock density, as measured by \citet{0004-637X-748-2-117}, of 2~cm$^{-3}$ with an increase by a factor of 4   assumed for the post-shock region. We estimate the `bullet' volume using our \textit{Chandra} image, assuming an ellipsoid with semi-minor and semi-major axes of 3$''$ and 5$''$, respectively. Combined with the velocity estimates listed in Table~\ref{tab:3}, this gives a possible $E_K$ range of 0.3 to 5$\times10^{49}$~erg, or a fraction to a few per cent of the total SN energy. Our mass estimate is likely an upper limit assuming a filling fraction less than 1 or a non-uniform density profile (i.e. not a monolithic `bullet'). The kinetic energy uncertainty budget is dominated by the poorly constrained velocity as it enters the calculation squared. Although large, our crude $E_K$ is similar to that estimated for the jet of Cas~A \citep{2003A&A...398.1021W,2016ApJ...822...22O}. On the other hand, this is more than measured for  `shrapnel' from the Vela \ac{SNR} \citep{2017A&A...604L...5G}.

What alternative scenario could explain the `bullet'? Its morphology could suggest a bow-shock \ac{PWN} origin. The main difference would be that a bow-shock \ac{PWN} would not be expected to have such strong Si and S lines, as seen in the X-ray spectrum \citep{0004-637X-748-2-117}. While a \ac{PWN} origin is unlikely, one way to distinguish such a scenario is through high-resolution and high-sensitivity radio observations that measure the change in the spectral index within this region. Specifically, a more flat spectral index ($-0.3<\alpha<0$) would favour a \ac{PWN} origin, while a steeper index ($\alpha<-0.5$) would point toward a \ac{SN} explosion origin. In Figure~\ref{figspectra} we can see some flattening of the spectral index ($-0.45<\alpha<-0.35$), compared to the rest of the remnant. But, this spectral index range falls into the ``grey'' area where neither scenario can be favoured.

Several galactic \ac{SNRs} are characterised by knot-like ejecta, and such clumps have been observed in both major types of SNe. Examples include the Tycho \ac{SNR} as type~Ia \citep{2017ApJ...834..124Y}, and Cas~A, G292.0+1.8, Puppis~A and big-Vela as core-collapse \citep{2017A&A...604L...5G} . These Galactic \ac{SNRs} exhibit a large range of ages -- from a few hundred to tens of thousands of years old. Knots from ejecta have been detected beyond the main shock in big-Vela \citep{2010ApJ...725.2038D} and Tycho \citep{2017ApJ...834..124Y}. Also, supersonic speeds in either bow-shock \ac{PWN} or `bullets' are not unusual, as is documented in several studies \citep[see their table~1]{2017JPlPh..83e6301K}. X-ray morphology and radio polarisation indicate a strong physical connection of this feature to the \ac{SNR} similar to many known Galactic \acp{SNR} such as Vela and Cas~A.

\subsection{Polarisation and Faraday Rotation}

Polarisation fraction and Faraday rotation provide powerful tracers of magnetic field orientation and strength.

\begin{figure*}
    \centering
    \includegraphics[width=0.45\textwidth,
                     %angle=-90,
                     trim=0 75 0 0, 
                     clip]{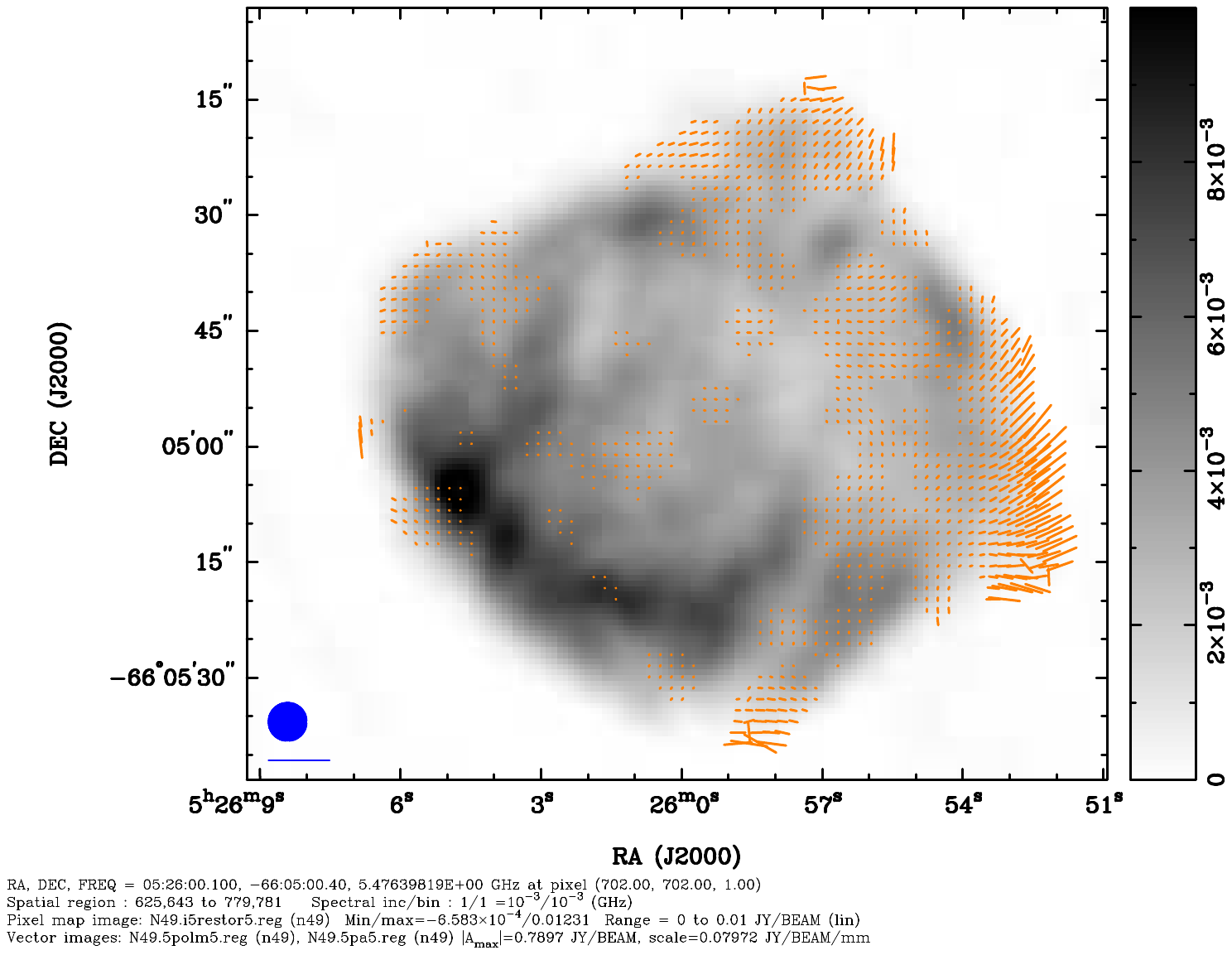}
    \includegraphics[width=0.45\textwidth,
                     %angle=-90,
                     trim=0 75 0 0,
                     clip]{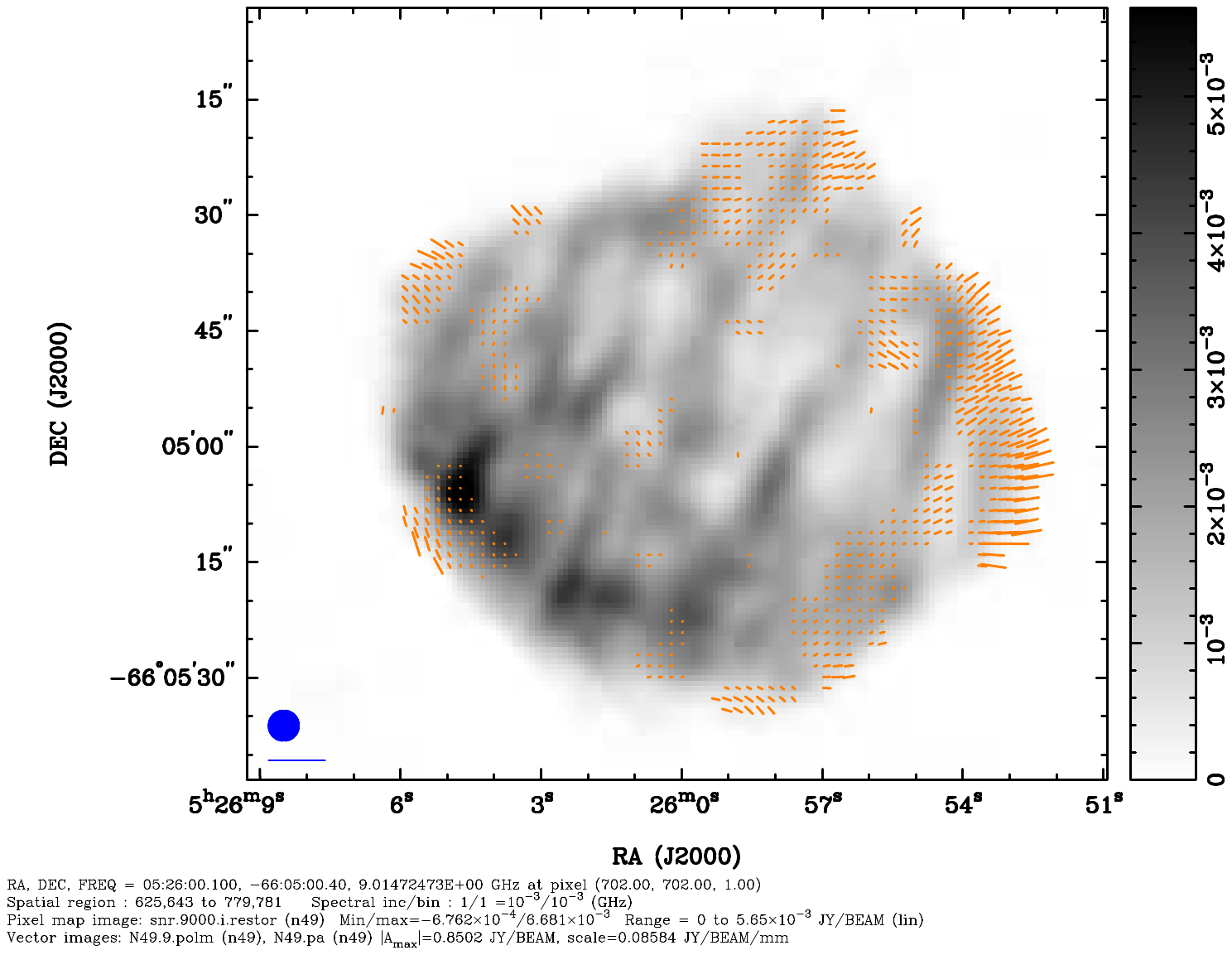}
 \includegraphics[width=0.45\textwidth,
                  %angle=-90,
                  trim=0 0 0 0, 
                  clip]{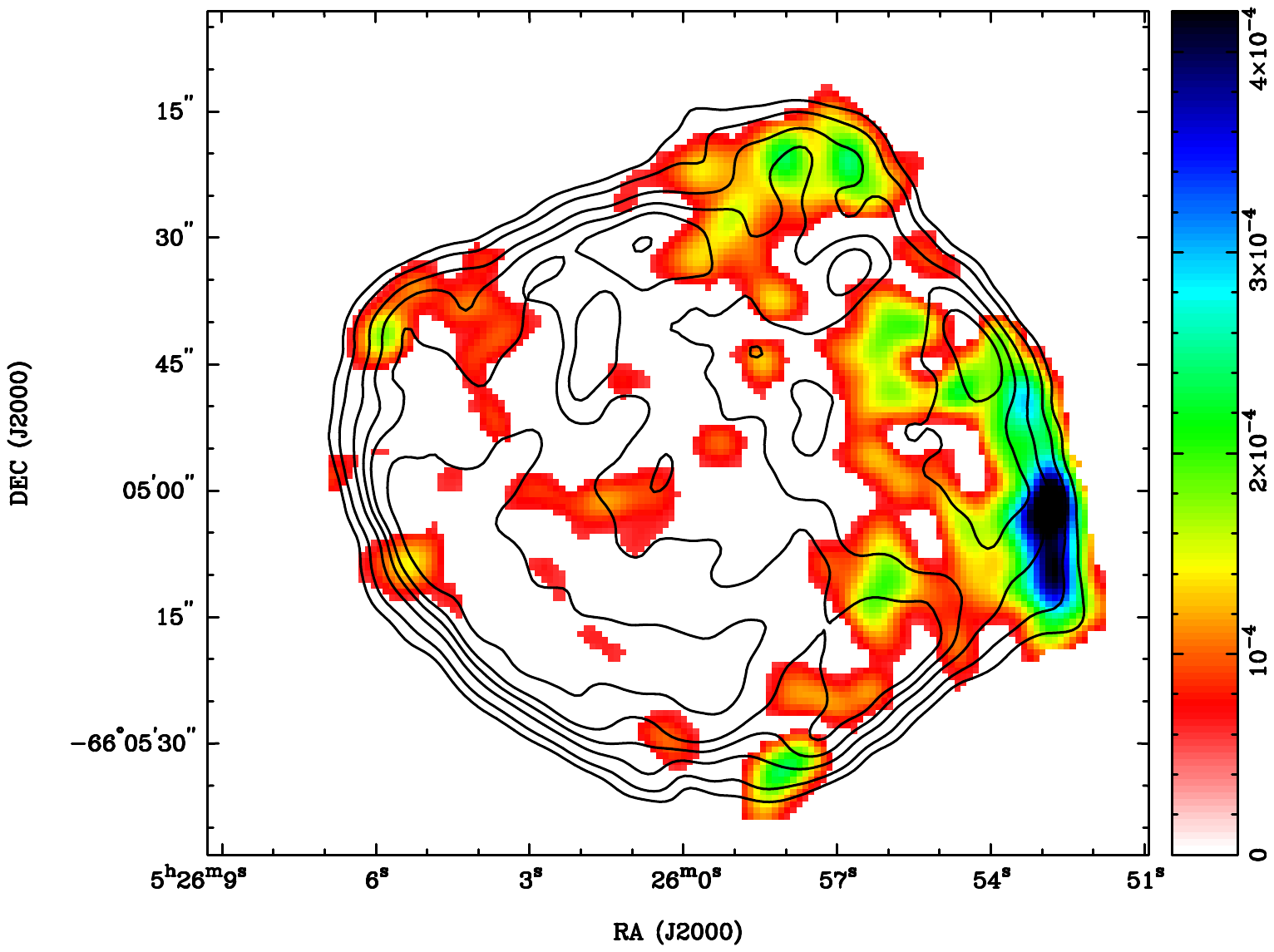}
    \includegraphics[width=0.45\textwidth,
                    %angle=-90,
                    trim=0 0 0 0, 
                    clip]{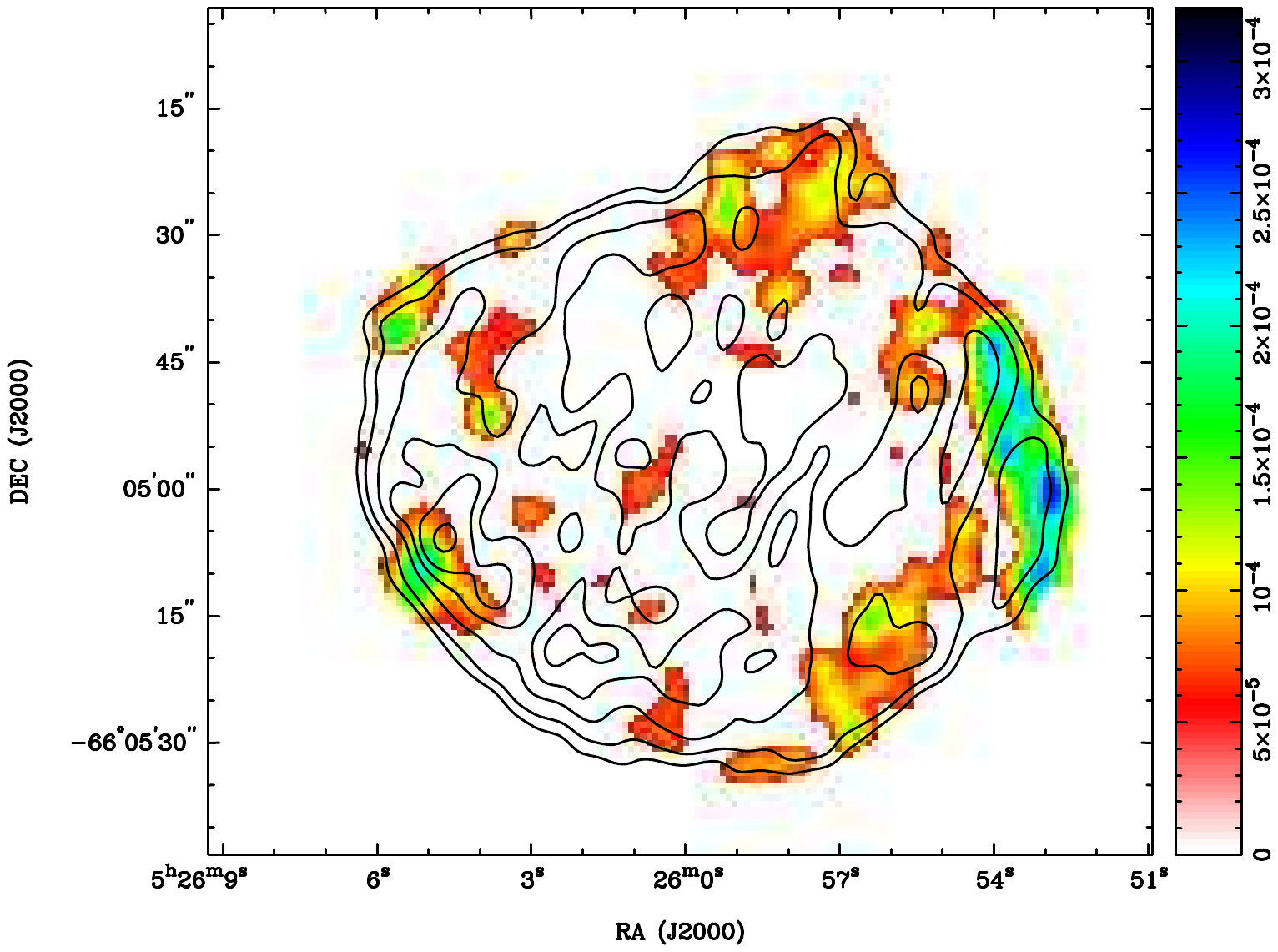}
    
    \caption{The polarisation map of \n49\ at 5.5~GHz (left column) and at 9~GHz (right column) where respective fractional polarisation vectors shown on the top images are overlaid on corresponding intensity \ac{ATCA} images. The blue circles in the lower left corner represent a synthesized beam of $5''\times~5''$ and $4''\times~4''$, for 5.5 and 9~GHz, respectively.  The blue line below the circles represents 100\%  polarization. The bar on the right side represents the grayscale intensity gradients for the \ac{ATCA} images in Jy\,beam$^{-1}$. Polarisation intensity maps at 5.5~GHz (bottom left) and at 9~GHz (bottom right) are shown in the bottom images with radio intensity contour lines overlaid; radio contours are 0.05, 1, 2, 4, and 6~mJy\,beam$^{-1}$ for 5.5 and 9~GHz, respectively.}

 \label{fig:5}
\end{figure*}

\begin{figure*}
    \centering 
    \includegraphics[width=0.45\textwidth,
                     %angle=-90,
                     trim=0 70 0 0, 
                     clip]{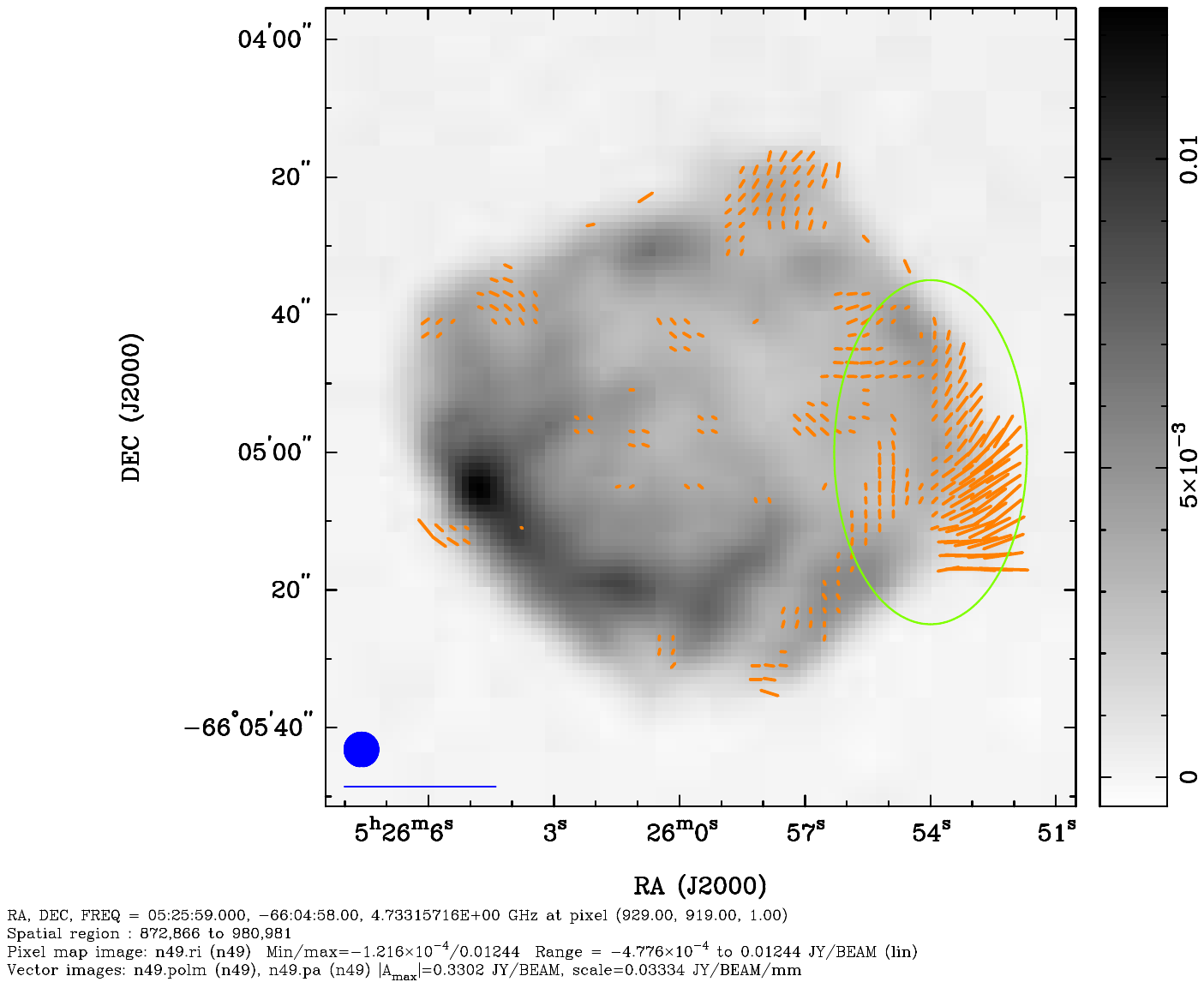}
    \includegraphics[width=0.45\textwidth,
                     %angle=-90,
                     trim=0 70 0  0, 
                     clip]{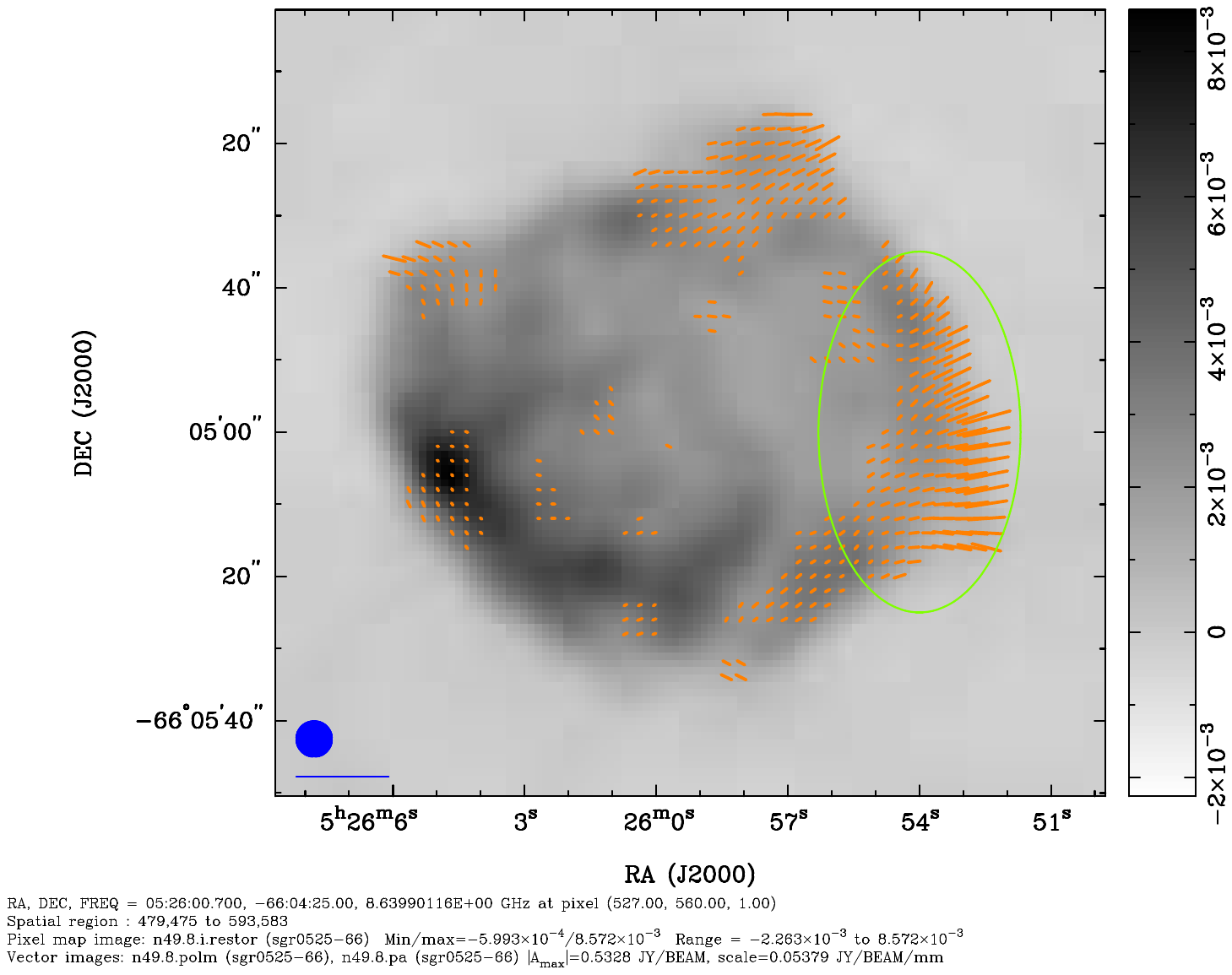} 
    
    \caption{
    Fractional polarisation vectors overlaid on their corresponding \ac{ATCA} intensity images.     
    The image on the left is at  4.79~GHz (pre-CABB).  large green ellipse in these images indicates an enhanced polarisation region around the proposed `bullet ejecta' or bow-shock \ac{PWN}.    
    The image on the right is at 8.64 GHz (pre-CABB). The blue circle in the lower left corner of this image represents the synthesized beam width of $5''\times~5''$ and the blue line below, a fractional polarisation vector of 100\%.}     
     
 \label{fig:6}
\end{figure*}

In Figure~\ref{fig:5} we show fractional polarisation vectors and polarisation intensity maps at 5.5, and 9~GHz. Fractional polarisation vectors for pre-CABB update images at 4.79 and 8.64~GHz are shown in Figure~\ref{fig:6}. We did not detect reliable polarisation emissions at 2.1 and 2.4~GHz. This is likely a result of depolarization induced by Faraday rotation, which is prominent at these frequencies.

%%%%%%%%%%%%%%%%%%%%%%%%%%%%%%%%%%%%%%%%%%%%%%%%%%%%%%%%%%%%%

%%%%%%%%%%%%%%%%%%%%%%%%%%%%%%% fig 5 %%%%%%%%%%%%%%%
%%%%%%%%%%%%%%%%%%%%%%%%%%%%%%%%%%%%%%%%%%%%%%%%%%%%%%%%

%%%%%%%%%%%%%%%%%%%%%%%%%%%% fig 6 %%%%%%%%%%%%%%%%%%

\begin{figure*}

\centering

\includegraphics[width=0.45\textwidth,]{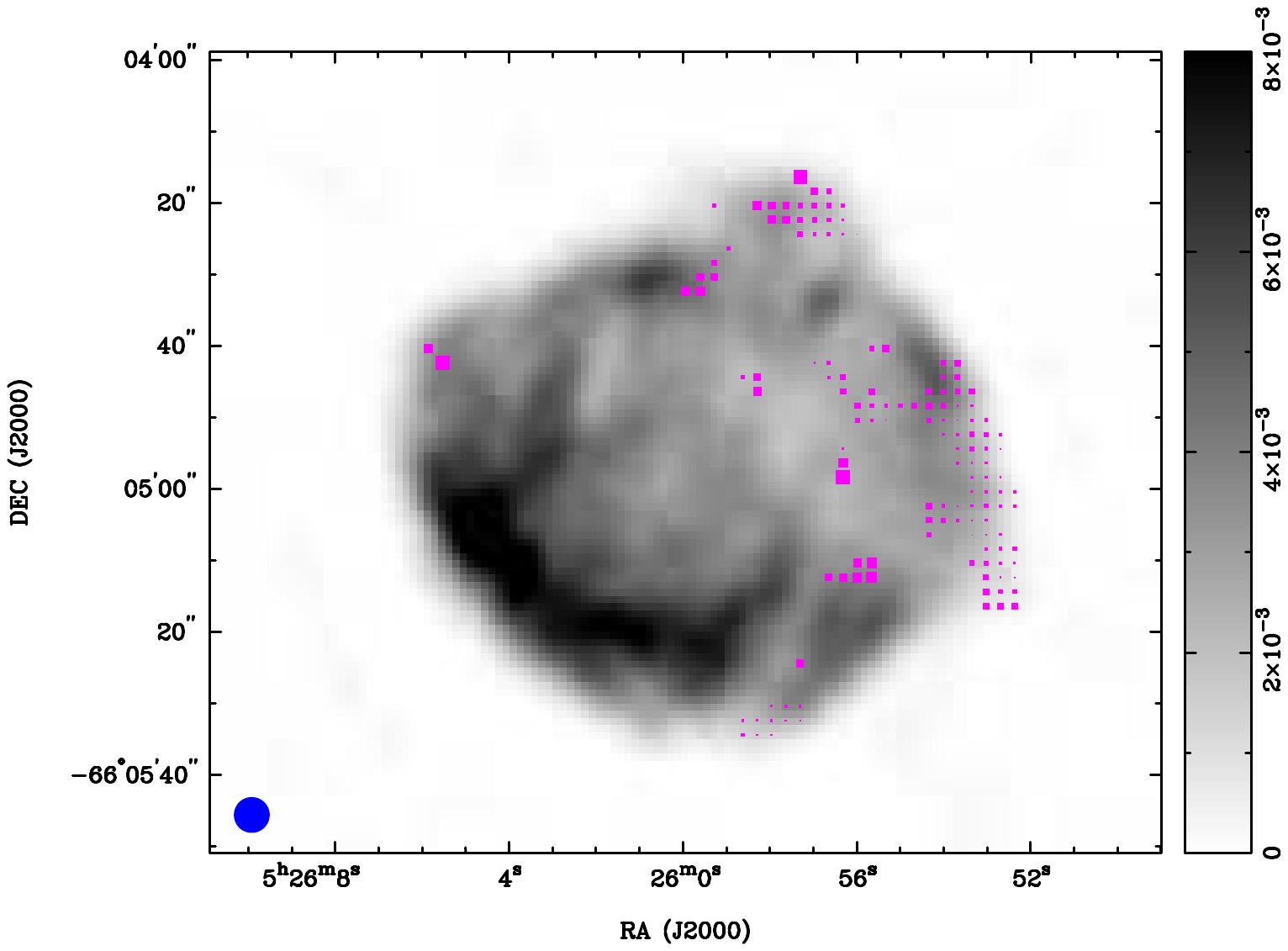}
    \includegraphics[width=0.45\textwidth,]{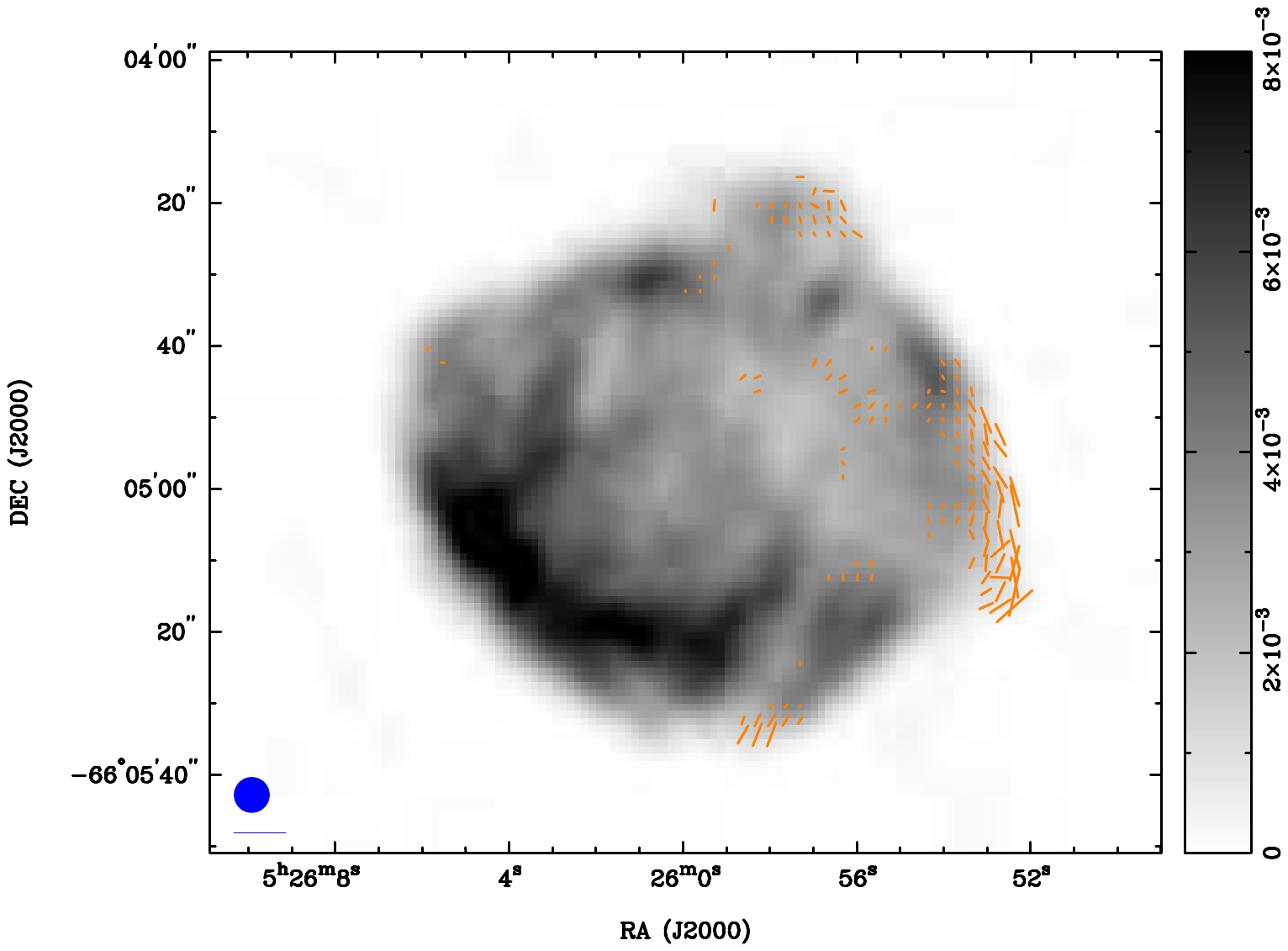} \caption{ The rotation measure map (left) of \n49\ overlaid on 5.5~GHz image. The pink boxes represent \ac{RM} values estimated from the position angles associated with linear polarisation. The maximum value of \ac{RM} is $\sim$591\,rad\,m$^{-2}$. The blue ellipse in the lower left corner represents the synthesised beam width of $5''\times~5''$.  The image on the right shows magnetic field vectors overlaying the 5.5 GHz radio image.
    }
     \label{fig:7}
\end{figure*}

We estimate a mean fractional polarisation at 5.5~GHz of 7$\pm$1\% with a maximum of 82$\pm$10\% (in the west). The mean fractional polarisation at 9~GHz is 10$\pm$1\% with a maximum of 85$\pm$13\%.  Pre-CABB data has significantly better {\it uv} coverage because 10 days of data were combined using different arrays. At 4.79~GHz we estimate a mean fractional polarisation of 8$\pm$3\% with a maximum of 33\%. The mean fractional polarisation at 8.64~GHz is even higher at 10$\pm$1\%  with a maximum of 47\%.  This is somewhat higher than a previous study of \n49\ by \citet{1998Dickel} who found a mean fractional polarisation across the remnant of  2.5\% at 4.79~GHz and 4.8\% at 8.64~GHz. This polarisation structure suggests a highly-organised magnetic field strongly aligned with the  `bullet' feature and its trail (see Figure~\ref{fig:7}).

Figure~\ref{fig:7} (left) illustrates the Faraday rotation of \n49\ which is determined from the position angles from the CABB data set as described above. The mean value for the Faraday rotation of \ac{SNR} \n49\ is 212 $\pm$ 65\, rad\,m$^{-2}$ and the maximum value of \ac{RM} is 591$\pm$103 rad\,m$^{-2}$. This \ac{RM} value is suggestive of a denser medium (as \ac{RM} is linear with $n_e$) as also evidenced by brightness in both radio and X-rays. The resulting magnetic field map shown in Figure~\ref{fig:7} (right) is a relative zero-level image obtained in the same way as for U and Q stokes. The magnetic field is stronger on the western side and coincides with the enhanced total intensity emission.

%jlp marker
%%%%%%%%%%%%%%%%%%%%%%%%%%%% fig 7 %%%%%%%%%%%%%%%%%%

%%%%%%%%%%%%%%%%%%%%%%%%%%%%%%%%%%%%%%%%%%%%%%%
We  use the equipartition formulae\footnote{http://poincare.matf.bg.ac.rs/$\sim$arbo/eqp/} \citep{Arbutina_2012,2013ApJ...777...31A,Urosevic2018} to estimate the magnetic field strength for \n49. The average equipartition field over the whole shell of \n49\ is $\sim$90\,$\mu$G with an estimated minimum energy\footnote{We use the following \n49\ values: $\theta$=0.625$'$, $\kappa=0$, S$_{\rm 1\, GHz}$=1.835~Jy and f=0.25; for $\kappa\neq0$ we estimate the average equipartition field of 231\,$\mu$G with an estimated minimum energy of E$_{\rm min}$=1.1$\times10^{50}$\, erg which is probably an overestimate because the physical background gives better equipartition arguments for $\kappa=0$ \citep{Urosevic2018}.} of E$_{\rm min}$=1.7$\times10^{48}$\, erg, assuming a power-law distribution for electrons, denoted in the model by `kappa', $\kappa$, equal to zero.

Although \n49\ is most likely a middle-aged \ac{SNR}, the equipartition assumption is not ideal for the determination of the magnetic field strength (perhaps only valid as an order of magnitude estimate, as used in \citet[][]{Urosevic2018}). In any case, the estimated value of $\sim$90\,$\mu$G can be explained by the magnetic field amplification at the strong shocks in relatively young to middle-aged and highly luminous \ac{SNRs}.

\subsection{\n49 Spectral Index}

Radio \ac{SNRs} often follow a power-law spectrum in which the spectral index, $\alpha$, is steep as defined by   $S_{\nu}$~$\propto$~$\nu^\alpha$; where   $S_{\nu}$ represents flux density at  $\nu$  frequency. To accurately measure the spectral index of \n49, we combined our observations with integrated flux density measurements across a wide range of frequencies obtained from the Murchison Widefield Array \citep[MWA; 84--200\, MHz; ][]{2018MNRAS.480.2743F}, Molonglo, Parkes, and \ac{ATCA} (408--14700\, MHz) radio telescopes, as shown in Table~\ref{tabflux}. 

\begin{figure*}
 \begin{center}
  \includegraphics[width=0.7\textwidth
,keepaspectratio]{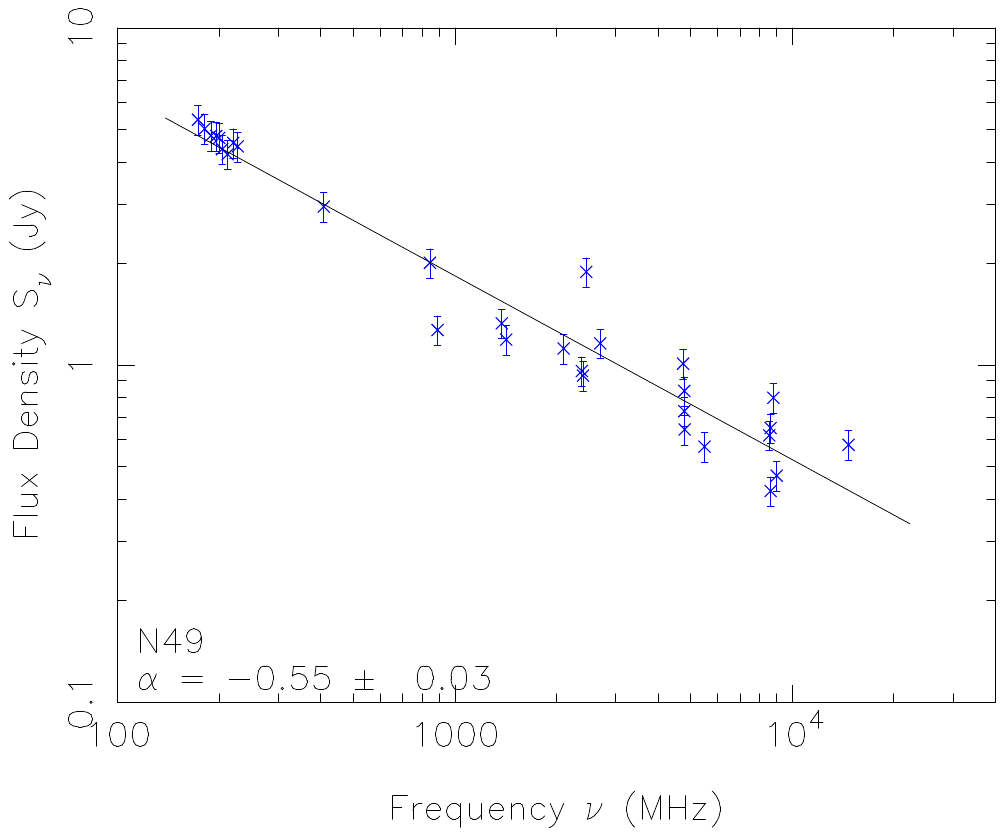} 
\end{center}
\caption{ Radio continuum spectrum of \n49 using MWA, ASKAP and \ac{ATCA} data (Table~\ref{tabflux}). The black solid line shows a linear least squares fit in logarithmic space, giving a spectral index of $\alpha= -0.55 \pm 0.03$. The relative errors (assumed to be 10\%)  are shown as vertical bars.}

\label{figspectra}

\end{figure*}

In Figure~\ref{figspectra}, we plot integrated flux densities vs.~frequency. Relative errors, assumed to be 10\%, are used as error bars on the logarithmic plot. The resulting radio spectral index, $\alpha= -0.55 \pm 0.03$, is typical of middle-age \ac{SNRs} \citep{1998A&AS..130..421F,2014Ap&SS.354..541U,2017ApJS..230....2B}.

%%%%%%%%%%%%%%%%%%%%%%%%%%%%%%%%%%%%%%%% FIG 6 new  %%%%%%%%%%%%%%%%%%%%%%%%%%%%
%%%%%%%%%%%%%%%%%%%%%%%%%%%%%%%%%%%%%%%%%%%%%%%%%%%%%%%%%%%%%%%%%%%%%%%%%%

\ac{SNR} \n49\ has an apparent diameter of 75$''$, which at the \ac{LMC} distance of 50~kpc corresponds to 18.2~pc. We estimate its flux density at 1~GHz to be S$_{\rm 1\,GHz}=1.82\pm0.04$~Jy, the surface brightness to be 1.75$\times10^{-19}$\,W\,m$^{-2}$\,Hz$^{-1}$\,sr$^{-1}$, and the total radio luminosity (10~MHz -- 100~GHz)  9.22$\times10^{27}$\,W. The position of  \n49\ within the surface brightness to diameter ($\Sigma$--\textit{D}) diagram ($\Sigma$=1.8$\times10^{-19}$\, W\,m$^{-2}$\, Hz$^{-1}$\,sr$^{-1}$, D=18.2\,pc) by \cite{2018ApJ...852...84P} suggests that this remnant is in its early-to-mid Sedov phase \citep{2020NatAs...4..910U,2022PASP..134f1001U}. This is consistent with an explosion energy of 2$\times10^{51}$\, erg evolving in an environment having a density of $\sim$0.02\,cm$^{-3}$. Such a low-density environment, however, is not consistent with our estimate of the remnant's  \ac{RM}. 
This is because a low-density medium would not provide enough electrons to produce significant Faraday rotation. Therefore, for an \ac{SNR} with a high \ac{RM}, the surrounding environment must have a relatively higher electron density, suggesting interaction with denser regions of the \ac{ISM}. This denser medium would contribute to the higher electron density required to explain the observed \ac{RM} values, however, \ac{RM} also depends on the strength of the magnetic field along the line of sight ($B_{||}$) where a stronger magnetic field could, in principle, increase the \ac{RM}.

\section{Discussion}
\label{disc}

\subsection{\n49\ Emission Features }

\begin{figure*}
\includegraphics[width=\textwidth,trim=0 0 5 0, clip]{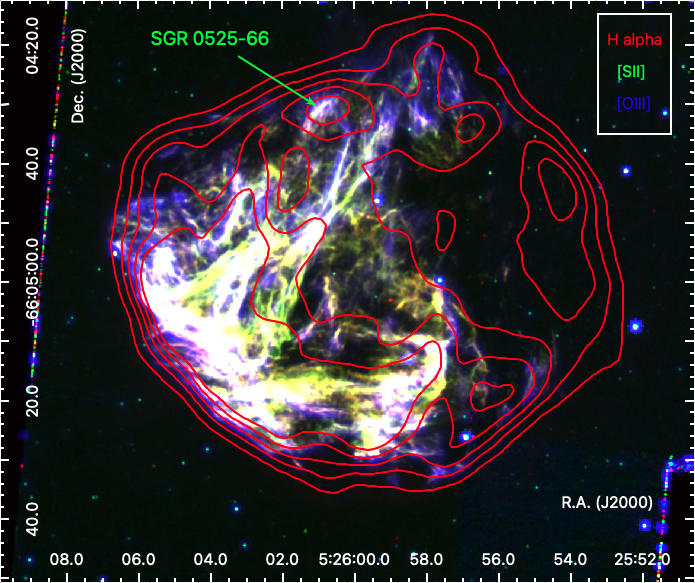}

\caption{ The HST optical image of \n49\ as described in \citet{2007AJ.Bilikova} (RGB = H$\alpha$ (red), \SII (green), and \OIII\ (blue)   overlaid with 5.5~GHz radio contours (0.25, 1.4, 2.6,3.8,5 and 5~mJy\,beam$^{-1}$). } 

     \label{fig:13}
\end{figure*}

\n49 is one of the brightest optical \acp{SNR} in the \ac{LMC}. While its semi-circled distinctive filamentary crescent opens to the northwest and emits very faint emission, a bright emission ridge can be seen in the east and southeast as shown in Figure~\ref{fig:13}.
Such filamentary structures in optical images indicate that 
optical filaments observed at distances of 5--10~pc may not be directly associated with the material that was initially surrounding the progenitor star but rather with the more general \ac{ISM} that the shockwave has encountered and interacted with as it expanded.
Massive progenitor stars usually clear most of their surrounding material during their lifetimes, due to their powerful stellar winds \citep{2007AJ.Bilikova}. 
 
\n49\ is located within a high column density \HI\ ridge between two kpc-scale super-giant shells, LMC4 and LMC5 \citep{1980MNRAS.192..365M}.  These are thought to have formed from multiple generations of stellar feedback \citep{2021MNRAS.505..459Fujii}. Multi-wavelength observations show clear evidence of interaction between the remnant and dense clumpy interstellar clouds on the eastern side  \citep{1992Vancura}. This interaction leads to bright emission in radio, optical, infrared, ultraviolet, and X-ray bands \citep{2003ApJ...586..210P,2007AJ.Bilikova}. The radio morphology of \n49 is generally consistent with trends seen at optical and X-ray wavelengths.

In the optical image, a line of brilliant filaments is seen as a ridge that runs south-east from the \ac{SNR}. We have overlaid 5.5~GHz radio contours on the optical image (Figure~\ref{fig:13}) to show the association between the optical and radio continuum emissions. We re-emphasise that \n49\ lacks bright filaments in its western part; this may be due to a cooler low-density environment of dust and gas. 

\begin{figure*}[ht!]
\centering
\includegraphics[ scale=0.6,trim=0 0 0 20,angle=-90,clip]{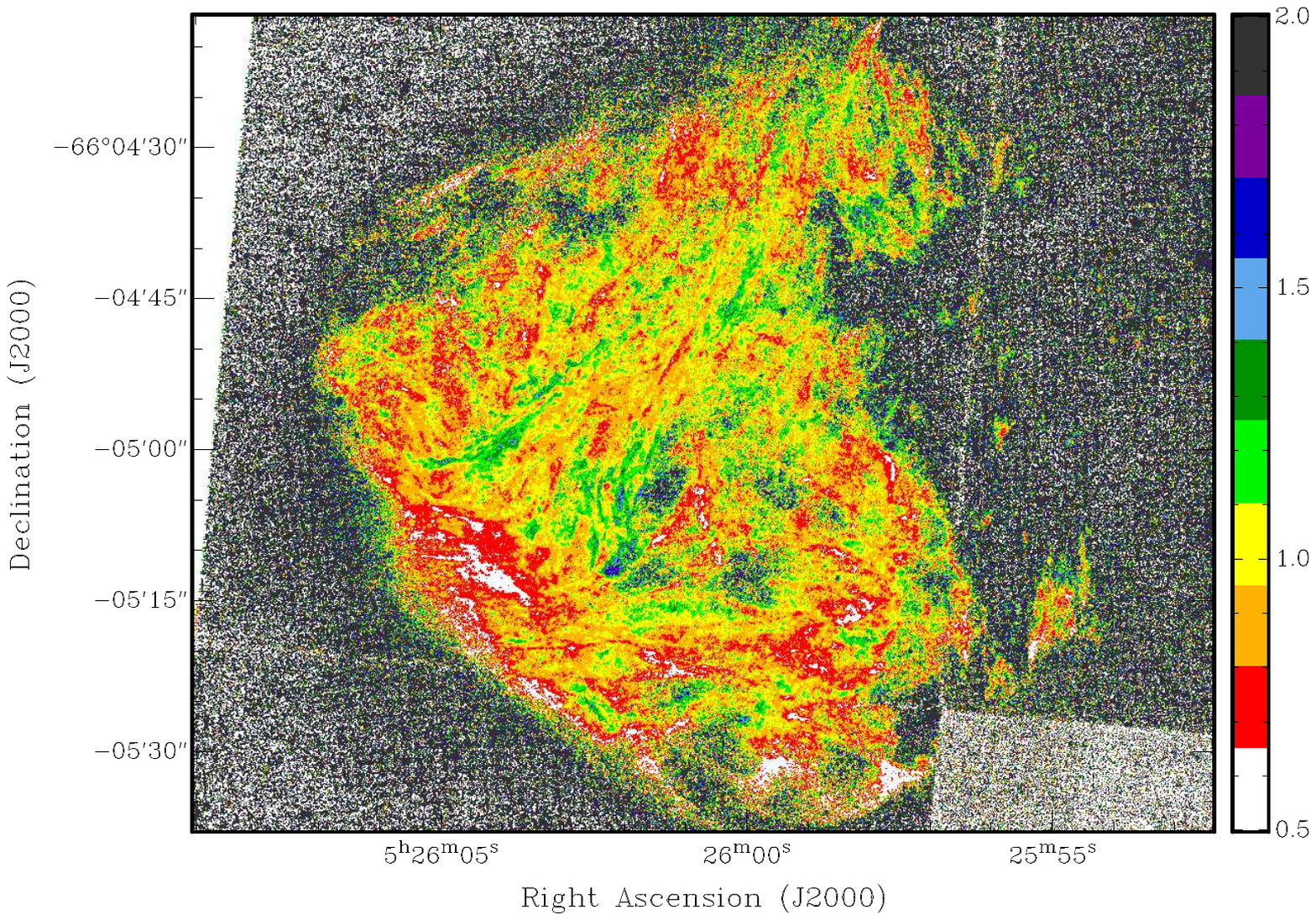}
\caption{The \SII/H$\alpha$ ratio map indicates the bright filaments and extent of emission.} 

     \label{fig:14}
\end{figure*}

The emission line ratio of singularly ionised sulfur to hydrogen-alpha (\SII/H$\alpha$ > 0.4) is primarily used in optical extragalactic searches for \acp{SNR}, as expected for shock-heated regions (Mathewson and Clarke 1973, D’Odorico et al. 1980, Fesen et al. 1984, Matonick and Fesen 1997, Blair and Long 1997). As it can be seen in Figure~\ref{fig:14}, \SII/H$\alpha$ ratios of 0.45--1.1 have been measured for this \ac{SNR} using HST images. The bullet feature shows only weak patches of H$\alpha$ emission on the west side of the remnant.   

\n49's X-ray/radio boundary of diffuse emission extends to the south and east along its circumference (Figures~\ref{fig:4}).  A diffuse emission component also appears to be embedded in the outermost emission patches which follow the X-ray/radio boundary.
 
X-ray and radio data confirm the cavity's presence in the remnant and asymmetric expansion of the gas has been observed in this region. This is associated with irregular density distributions of materials surrounding the explosion \citep{2022MNRAS.515.1676Campana}.

%%%%%%%%%%%%%%%%%%fig13%%%%%%%%%%%%%%%%%

%%%%%%%%%%%%%%%%%%%%%%%%%%%%%%%%%%%%%%%%%%%%%%%%%%%%%%%%%%%%%%%%%%%%%

%%%%%%%%%%%%%%%%%%%%%%%%%%%%%%%%%%%%

%%%%%%%%%%%%%%%%%%%%%%%%%%%%%%%%%%

%%%%%%%%%%%%%%%%%%%%%%%%% Table 4 %%%%%%%%%%%%%%%%%%%%%%%%%%%%%%%%
\begin{table}
\footnotesize
\caption{Flux density measurements of \ac{SNR} \n49\ at multiple radio frequencies. $\dag$ indicates that we used images from \citet{2018MNRAS.480.2743F} to measure \n49\ flux densities.}
\label{tabflux}
\begin{tabular}{@{}cccl@{}}
\hline\hline
$\nu$ & Flux Density & Telescope & Reference\\
(MHz) &$S_{\nu}$(Jy)\\
\hline
173$\dag$&5.34&MWA&This work\\
181$\dag$&5.02&MWA&This work\\
189$\dag$&4.79&MWA&This work\\
196$\dag$&4.78&MWA&This work\\
200$\dag$&4.72&MWA&This work\\
204$\dag$&4.37&MWA&This work\\
212$\dag$&4.22&MWA&This work\\
220$\dag$&4.56&MWA&This work\\
227$\dag$&4.45&MWA&This work\\
408&2.95&Molonglo&\citet{2017ApJS..230....2B}\\
843&2.01&MOST&\citet{2017ApJS..230....2B}\\
888&1.27&ASKAP&\citet{2021MNRAS.506.3540P}\\
1377&1.33&ATCA&\citet{2017ApJS..230....2B}\\
1420&1.19&ATCA&\citet{2017ApJS..230....2B}\\
2100&1.09&ATCA&This work\\
2378&0.960&ATCA&\citet{2017ApJS..230....2B}\\
2400&0.930&ATCA& This work \\
2450&1.890&ATCA&\citet{2017ApJS..230....2B}\\
2700&1.16&ATCA&\citet{2017ApJS..230....2B}\\
4790&0.838&ATCA& This work \\
4750&1.01&Parkes&\citet{2017ApJS..230....2B}\\
4790&0.730&Parkes&\citet{2017ApJS..230....2B}\\
4800&0.644&ATCA&\citet{2017ApJS..230....2B}\\
5500&1.02&ATCA&This work\\
8550&0.619&Parkes&\citet{2017ApJS..230....2B}\\
8640&0.651&ATCA& This work \\
8640&0.423&ATCA&\citet{2017ApJS..230....2B}\\
8800&0.800&Parkes&\citet{2017ApJS..230....2B}\\
9000&0.30&ATCA&This work\\
14700&0.58&Parkes&\citet{2017ApJS..230....2B}\\
\hline
\end{tabular}
\end{table}
%%%%%%%%%%%%%%%%%%%%%%%%%%%%%%%%%%%%%%%%%%%%%%%%%%%%%%%%%%%%%%%%%%%%%

\subsection{Comparison with other similar SNRs}
 \label{comp}

 Studies of \ac{SNRs}, including \n49, are essential because they provide unique constraints on the properties of these systems \citep{2008Gaensler}. We compare \n49 with several well-studied \ac{SNRs} and \ac{PWN}e in the \ac{MCs} and the Milky Way Galaxy. For example, both  MCSNR~J0540--693 and MCSNR~J0453--6829 in the \ac{LMC} have \ac{PWN}e; the former hosting a 50~ms pulsar \citep{2000Gotthelf}, while the latter has an estimated spectral index   $\alpha=-0.39\pm0.03$, consistent with non-thermal radio shell emission  \citep{2012Harbel}.  N\,206 (B\,0532--71.0) is an \ac{SNR} in the \ac{LMC}  suggested to be in the middle to late stage of its evolution \citep{2002AJ....124.2135K} with a small X-ray and radio continuum knot at the outer tip. This feature's elongated morphology and the surrounding wedge-shaped X-ray enhancement suggest a possible bow shock \ac{PWN} structure, with a spectral index over the remnant of --0.33. This value is fairly flat for an \ac{SNR}; spectral indices for  \ac{PWN}e tend to be greater than --0.4, as compared to a typical shell \ac{SNR} that tends to be between --0.8 and --0.3. An X-ray study of N\,206 suggests this \ac{SNR}-\ac{PWN} system is generated by an eastward moving pulsar \citep{2005Williams}.

We observe a much wider range of \ac{PWN}e and \acp{SNR} in our Galaxy. For example, IC\,443 is a 
mixed-morphology\footnote{Known as thermal composite \acp{SNR}, these appear as a regular radio shell-type remnant with centrally-peaked thermal X-ray emission.} Galactic \ac{SNR} is notable for its interaction with surrounding molecular clouds. The \ac{PWN} is detected in the southern edge of the \ac{SNR} and is suggested to be associated; it was produced around 30\,000~yrs ago in a core-collapse SN event \citep{2001Olbert}. Observational characteristics of this \ac{PWN} show a flat spectrum radio emission ($\alpha$=~--0.1 to --0.3), a steep X-ray spectrum ($\alpha_{x}$=~1.0~--~1.5) and a high degree of linear polarisation ($>$5\%). The tail of the nebula does not point toward the geometric centre of the \ac{SNR} \citep{2001Olbert}. The pulsar B\,1951+32A is associated with a bullet-shaped \ac{PWN} having a bow-shocked morphology, seen in radio and X-ray as moving with a supersonic velocity. This pulsar is associated with the strongly polarised ($\sim$15--30\%)  non-thermal Galactic radio \ac{SNR} CTB\,80 \citep{2003Castelletti}. The radio spectral index of this \ac{SNR} has been estimated to be  $\alpha=-0.36\pm0.02$ \citep{2005Castelletti,1970PhDT.Downes}. 
 
\n49\ contrasts with IC\,443, N\,206 (MCSNR J0531--7100) and CTB\,80 which show similarities in radio spectral index, radio emission, and morphology.  \n49s   typical \ac{SNR} radio spectral index  ($\alpha=-0.54\pm0.02$) and its high radio polarisation in the southwestern part suggests a knot scenario is more probable than a \ac{PWN} explanation.

\section{Conclusions and future studies}
 \label{conclusion}

We present new radio continuum intensity and polarisation images of the well-known \ac{LMC} \ac{SNR} \n49. A high fractional polarisation region is found around the area where the proposed `bullet' ejecta is positioned with moderate levels of mean fractional polarisation at 5.5 and 9~GHz of 7$\pm$1\% and 10$\pm$1\%, respectively, throughout the remnant as a whole. In the vicinity of the so-called `bullet' knot, these polarisation values are noticeably larger than found in previous studies.

The mean value for the Faraday rotation of \ac{SNR} \n49 from combining  CABB data is 212$\pm$65~rad\,m$^{-2}$ and the maximum value of RM is 591$\pm$103~rad\,m$^{-2}$. 
The strong \ac{RM} near the southwest of the \ac{SNR} might correspond with strong hard X-rays positioned to the southeast. 

Together with {\it Chandra} X-ray observations, our analyses of new and archival (pre-CABB; 1990s) \ac{ATCA} observations 
has allowed further study of prominent features of the  `bullet' found beyond the southwestern boundary. This presumed ejecta knot is possibly interacting with a molecular cloud, causing increased brightness in this region while no polarisation can be seen in the cloud proper. Thermal soft X-ray emissions may barely be seen within this region in the three-colour image of \n49\ (Figure~\ref{fig:4}).

We note the explosion centre of \n49, based on the knot's trail projection is offseted from the geometrical centre of the \ac{SNR} ($\sim$44$''$ or $\sim$10.4~pc).

The next generation's high dynamic range and resolution radio observations of \n49\ are essential for establishing the connection and nature of the knot's polarisation, magnetic field, and Faraday rotation.

\begin{acknowledgement}
The \ac{ATCA} is part of the Australia Telescope National Facility which is funded by the Australian Government for operation as a National Facility managed by \ac{CSIRO}.
This paper includes archived data obtained through the Australia Telescope Online Archive (http://atoa.atnf.csiro.au). 
We used the \textsc{karma} and \textsc{miriad} software packages developed by the \ac{ATNF}. 
MDF and GR acknowledge ARC funding through grant DP200100784.
D.U. acknowledges the Ministry of Education, Science and Technological Development of the Republic of Serbia through contract No. 451-03-68/2022-14/200104, and for the support through the joint project of the Serbian Academy of Sciences and Arts and Bulgarian Academy of Sciences on the detection of extragalactic \acp{SNR} and \HII\ regions. We thank Jane Kaczmarek for her suggestions and support.
We thank an anonymous referee for comments and suggestions that greatly improved our paper.
\end{acknowledgement}

\bibliography{R2}

\end{document}